\documentclass[usenatbib]{mn2e}

\usepackage{graphicx}
\usepackage{subfigure}
\usepackage{subfloat}
\usepackage{amssymb}
\usepackage{amsmath}
\usepackage{multirow}
\usepackage{dcolumn}
\usepackage{bm}

\newcommand\abl{{\langle{\bf A}\cdot{\bf B}\rangle}}

\newcommand\bfv{{\bf v}}

\newcommand\beq{\begin{equation}}
\newcommand\eeq{\end{equation}}

\title{Influence of initial conditions on the large-scale dynamo growth rate}
\author{Kiwan Park$^{1}$\thanks{E-mail: pkiwan@pas.rochester.edu}\\
$^{1}$Department of Physics and Astronomy, University of Rochester, Rochester, NY 14627, USA}

\begin{document}
\date{}
\maketitle
\label{firstpage}

\begin{abstract}
To investigate the effect of energy and helicity on the growth of magnetic field, helical kinetic forcing was applied to the magnetohydrodynamic(MHD) system that had a specific distribution of energy and helicity as initial conditions. Simulation results show the saturation of a system is not influenced by the initial conditions, but the growth rate of large scale magnetic field is proportionally dependent on the initial large scale magnetic energy and helicity. It is already known that the helical component of small scale magnetic field(i.e., current helicity $\langle {\bf j}\cdot {\bf b}\rangle$) quenches the growth of large scale magnetic field. However, $\langle {\bf j}\cdot {\bf b}\rangle$ can also boost the growth of large scale magnetic field by changing its sign and magnitude. In addition, simulation shows the nonhelical magnetic field can suppress the velocity field through Lorentz force. Comparison of the profiles of evolving  magnetic and kinetic energy indicates that kinetic energy migrates backward when the external energy flows into the three dimensional MHD system, which means the velocity field may play a preceding role in the very early MHD dynamo stage.
\end{abstract}

\section{Introduction}
The generation and amplification of magnetic field in astrophysical systems are ubiquitous phenomena. The origin and exact mechanism of growth of magnetic fields in stars or galaxies have been long standing problems. It has been thought that helical kinetic motion or turbulence amplifies the magnetic field($\bf B$ field). However, the helical component does not seem to be an absolute necessity for the amplification of large scale magnetic field. In astrophysical dynamos, for instance, the kinetic energy of some celestial objects like supernovae or galaxy clusters has low or practically zero level of helical component. The evolution of $\bf B$ fields in these objects is thought to be dominated by small scale dynamo($SSD$): the amplification of fields below the large scale eddy without helicity(\cite{1968JETP...26.1031K}, \cite{1992ApJ...396..606K}, \cite{1981PhRvL..47.1060M}, \cite{2003ApJ...597L.141H}, \cite{2004ApJ...612..276S}), \cite{2005ApJ...626..853M}). So, it is important to understand the detailed mechanism of dynamo in MHD equations whether or not the driving force is helical.\\

\noindent As of yet some problems in the MHD dynamo process are not completely understood: the role of helical or nonhelical kinetic(magnetic) field, the effects of initial conditions($IC$s) such as kinetic(magnetic) energy and helicity. There were trials to see the effects of $IC$s on the dynamo(\cite{2004PhRvE..70c6408H}, \cite{2004ApJ...603..569M}). However, the trials are not yet enough; moreover, there are few analytic studies to explain the effects of initial conditions. Some statistical methods like Eddy Damped Quasi Normal Markovian approximation($EDQNM$, \cite{1976JFM....77..321P}) can be used to explain the influences of $IC$s on the profile of growing $\bf B$ field qualitatively. However, it is partial and incomplete. Development and verification of the theoretical results with more detailed simulation data are necessary. Nonetheless, related simulation results still provide us many detailed phenomena that are helpful to understanding the MHD turbulence. In this paper the effects of initial magnetic energy and helicity on the large scale dynamo were investigated using simulation data and analytic methods.

\section{Problem to be solved and methods}
The main aim of this paper is to find out the effect of initial conditions($IC$s) on the growth and saturation of magnetic helicity($H_M=1/2\langle {\bf A}\cdot {\bf B}\rangle$, ${\bf B}=\nabla \times{\bf A}$) and magnetic energy($E_M$). For this, the combinations of three simulations were carried out: Non Helical Magnetic Forcing($NHMF$), Helical Magnetic Forcing($HMF$), and Helical Kinetic Forcing($HKF$). To explain simulation results, the equations derived from $EDQNM$ and two scale mean field dynamo theory(\cite{2002ApJ...572..685F}) were used.
\\

\noindent For the simulation code, high order finite difference Pencil Code(\cite{2001ApJ...550..824B}) and the message passing interface(MPI) were used.
The equations solved for $HKF$ in the code are,
 \begin{eqnarray}
\frac{D \rho}{Dt}&=&-\rho \nabla \cdot {\bf v}\\
\frac{D {\bf v}}{Dt}&=&-c_s^2\nabla \mathrm{ln}\, \rho + \frac{{\bf J}\times {\bf B}}{\rho}+\nu\big(\nabla^2 {\bf v}+\frac{1}{3}\nabla \nabla \cdot {\bf v}\big)+{\bf f}\\
\frac{\partial {\bf A}}{\partial t}&=&{\bf v}\times {\bf B} -\eta\,{\bf J}
\label{MHD equations in the code}
\end{eqnarray}
$\rho$: density; $\bf v$: velocity; $\bf B$: magnetic field; $\bf A$: vector potential; ${\bf J}$: current density;  $D/Dt(=\partial / \partial t + {\bf v} \cdot \nabla$): advective derivative; $\eta$: magnetic diffusivity; $\nu$(=$\mu/\rho$, $\mu$: viscosity, $\rho$: density): kinematic viscosity; $c_s$: sound speed; $\bf f$: forcing function(helical or nonhelical). The unit used in the code is `$cgs$'. Velocities are expressed in units of $c_s$, and magnetic fields in units of $(\rho_0\,\mu_0)^{1/2}c_s$ (i.e., $B=\sqrt{\rho_0\,\mu_0}v$, $\mu_0$ is magnetic permeability and $\rho_0$ is the initial density). Note that $\rho_0\sim \rho$ in the weakly compressible simulations. These constants $c_s$, $\mu_0$, and $\rho_0$ are set to be `1'. In the simulations $\eta$(=$c^2/4\pi \sigma$, $\sigma$: conductivity) and $\nu$ are 0.006.
\\

\noindent In case of the magnetically driven simulation(magnetic forcing, $MF$), forcing function ${\bf f}$ is located in the magnetic induction equation($\partial {\bf A}/\partial t={\bf v}\times {\bf B} -\eta\,{\bf J}+{\bf f}$) instead of the momentum equation. As Ohm's law($\eta {\bf J} = {\bf E}+{\bf v}\times {\bf B}$) implies, ${\bf f}$ symbolizes sort of the external electromagnetic force that drives the magnetic eddy(\cite{1999PhPl....6.4146E}, \cite{2012MNRAS.423.2120P}).
\\

\noindent We employ a cube like periodic box of spatial volume $(2 \pi)^3$ with mesh size of $256^3$ for runs. The forcing function ${\bf f}$(http://pencil-code.nordita.org) used in the simulations is either fully helical(in fourier space, $\nabla\times{\bf f}=k\,{\bf f}$, $k$: wave number) or non-helical($\nabla\times{\bf f}\neq k\,{\bf f}$). ${\bf f}(x,t)$ is represented by $N\,{\bf f_k}(t)\, exp\,[i\,{\bf k}_\mathrm{f}(t)\cdot {\bf x}+i\phi(t)]$($N$: normalization factor, ${\bf k}_\mathrm{f}(t)$: forcing wave number). And to prevent the shock phenomenon, forcing magnitude ${\bf f_k}$ is $0.07$ for $KF$ and 0.01 for $MF$(note that $\nabla \times {\bf f}=k_f {\bf f}$ for the helical forcing). This makes mach number(=$v/c_s$) less than 0.3.

\section{Simulation Result 1}
\begin{figure}
\centering{
{%
   \subfigure[$|H_M|$ and $E_M$ (Logarithmic scale)]{
     \includegraphics[width=8 cm]{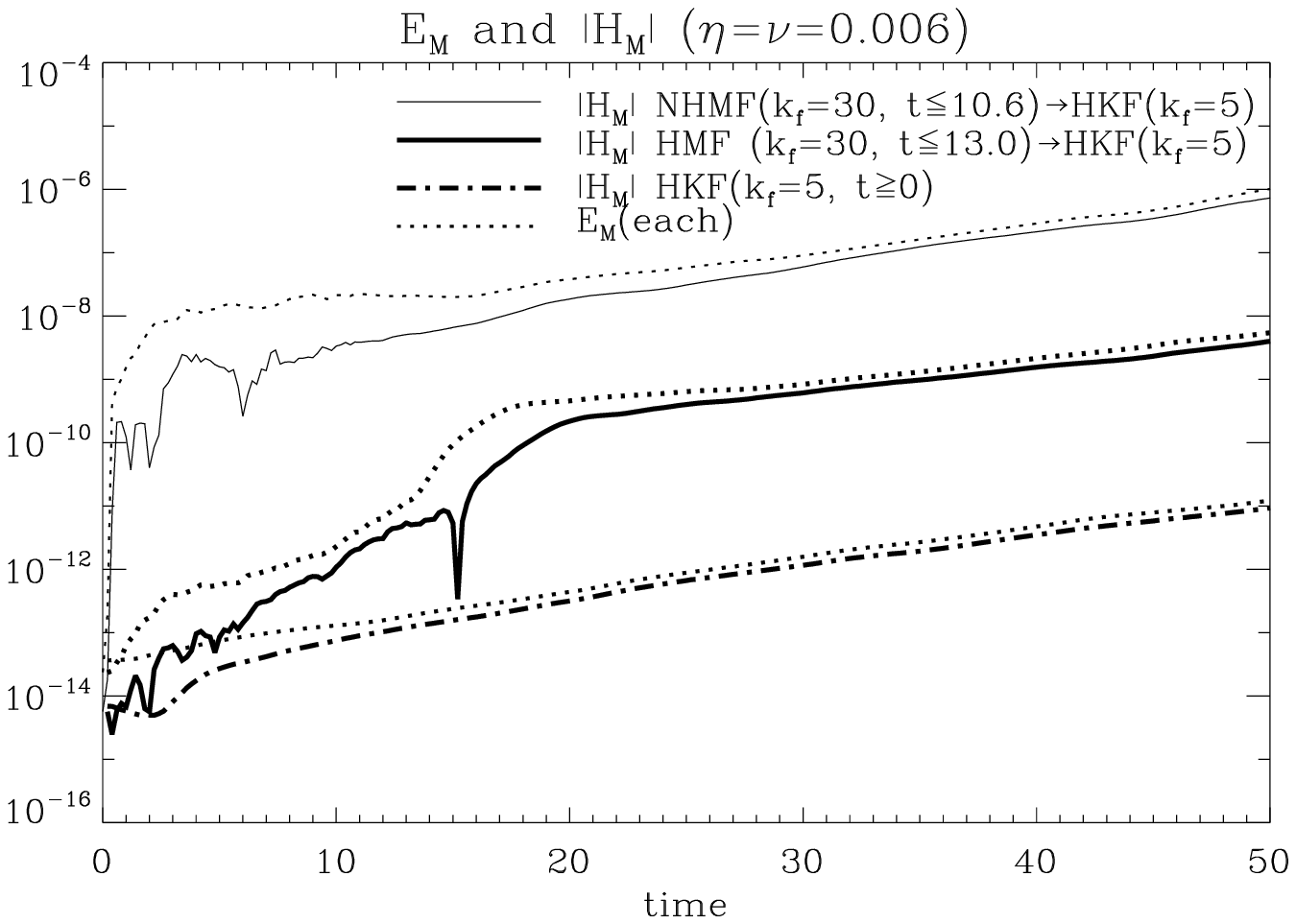}
     \label{1}}\,
   \subfigure[$H_M$ and $E_M$ (Linear scale)]{
     \includegraphics[width=8 cm]{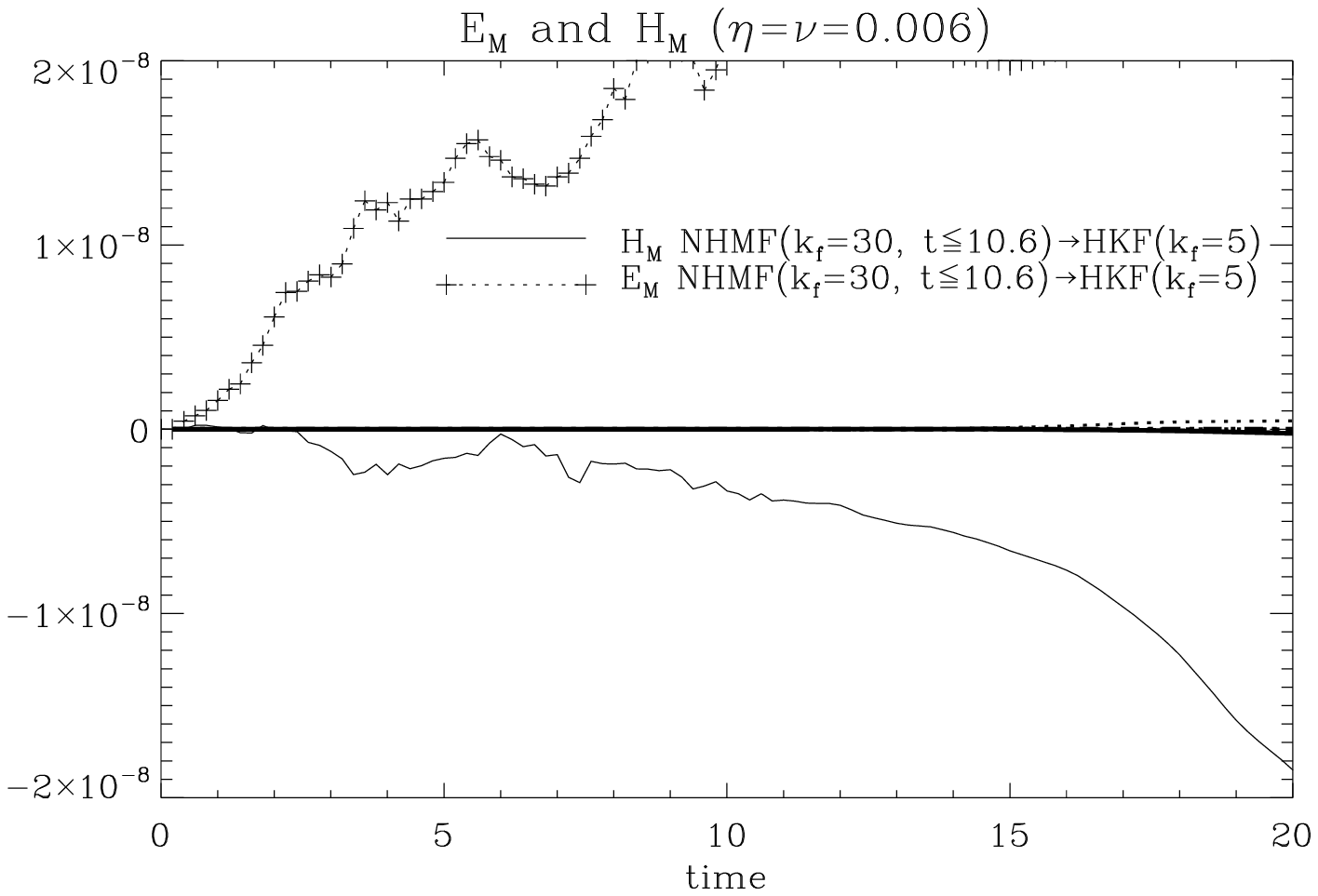}
     \label{2}
   }\,
   \subfigure[$|H_M|$ and $E_M$ (Linear scale)]{
     \includegraphics[width=8 cm]{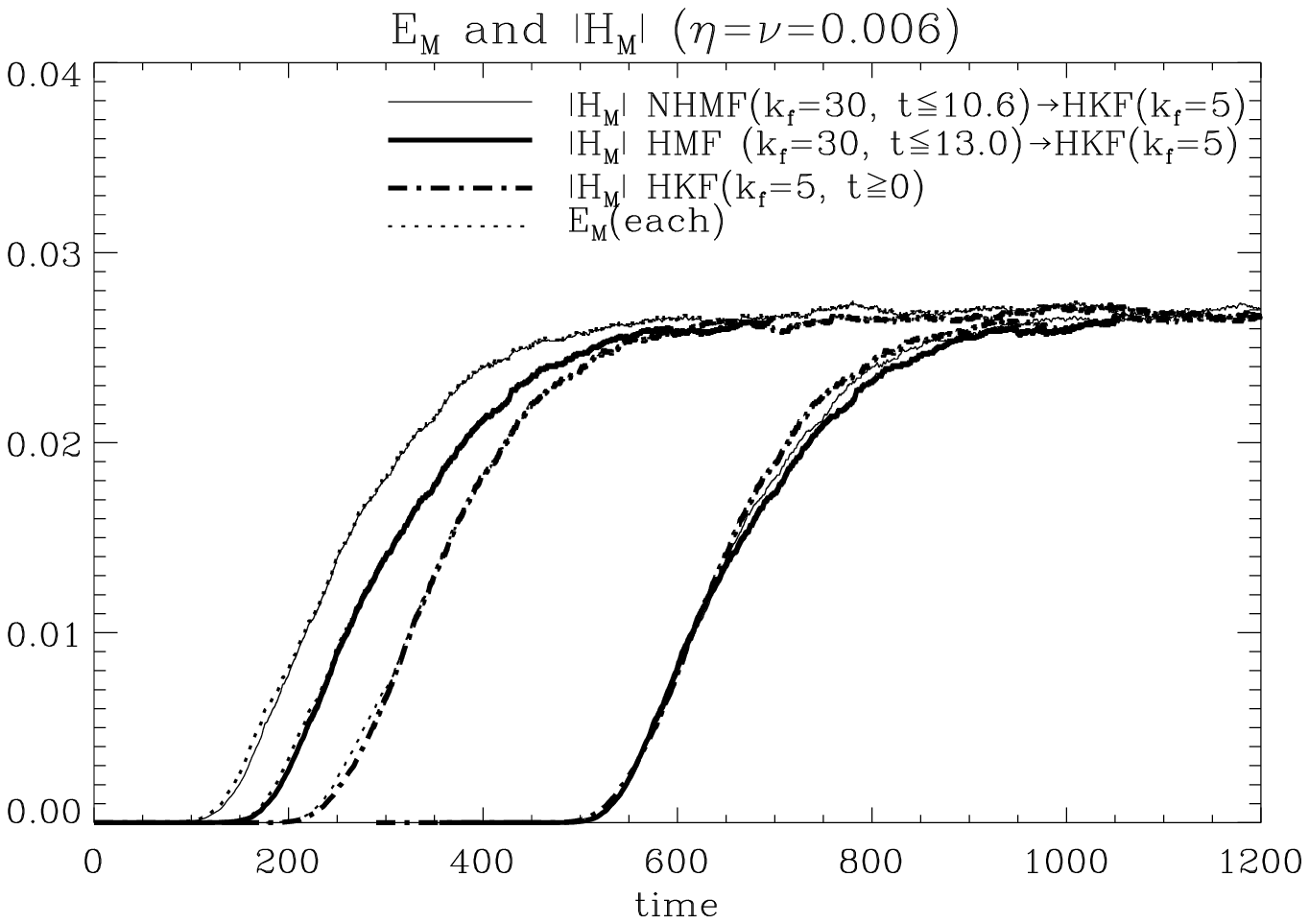}
     \label{3}
   }\,
     }
     }
\caption{Seed energy and helicity in each case are the same. Precursor simulation $(N)HMF$ changes the given seed field into the specific energy distribution, which is used as new initial conditions for the consecutive main simulation $HKF$. (a) Preliminary simulation $NHMF$($|f_k|=0.01$ at $k_f=30$) finishes at $t=10.6$. During this time regime, $H_M$ is negative. In contrast, $HMF$($|f_k|=0.01$ at $k_f=30$ for $t\leq13.0$) generates positive $H_M$. $HKF$($|f_k|=0.07$ at $k_f=5$) follows these preliminary simulations. And $HKF$ without a precursor simulation was done separately as a reference.(b) Except $NHMF$, the magnetic fields in the other cases are indistinguishably small. (c) The left line group shows the influence of $IC$s. The difference in the onset positions is mainly decided by large scale $E_{M0}$ and $H_{M0}$ generated by the precursor simulations. And right line group includes the shifted $E_M$ and $H_M$ of each case for the comparison.}
\end{figure}

\begin{figure}

{
{%
   \subfigure[$E_{kin}$ ($NHMF\rightarrow HKF$)]{
     \includegraphics[width=8cm]{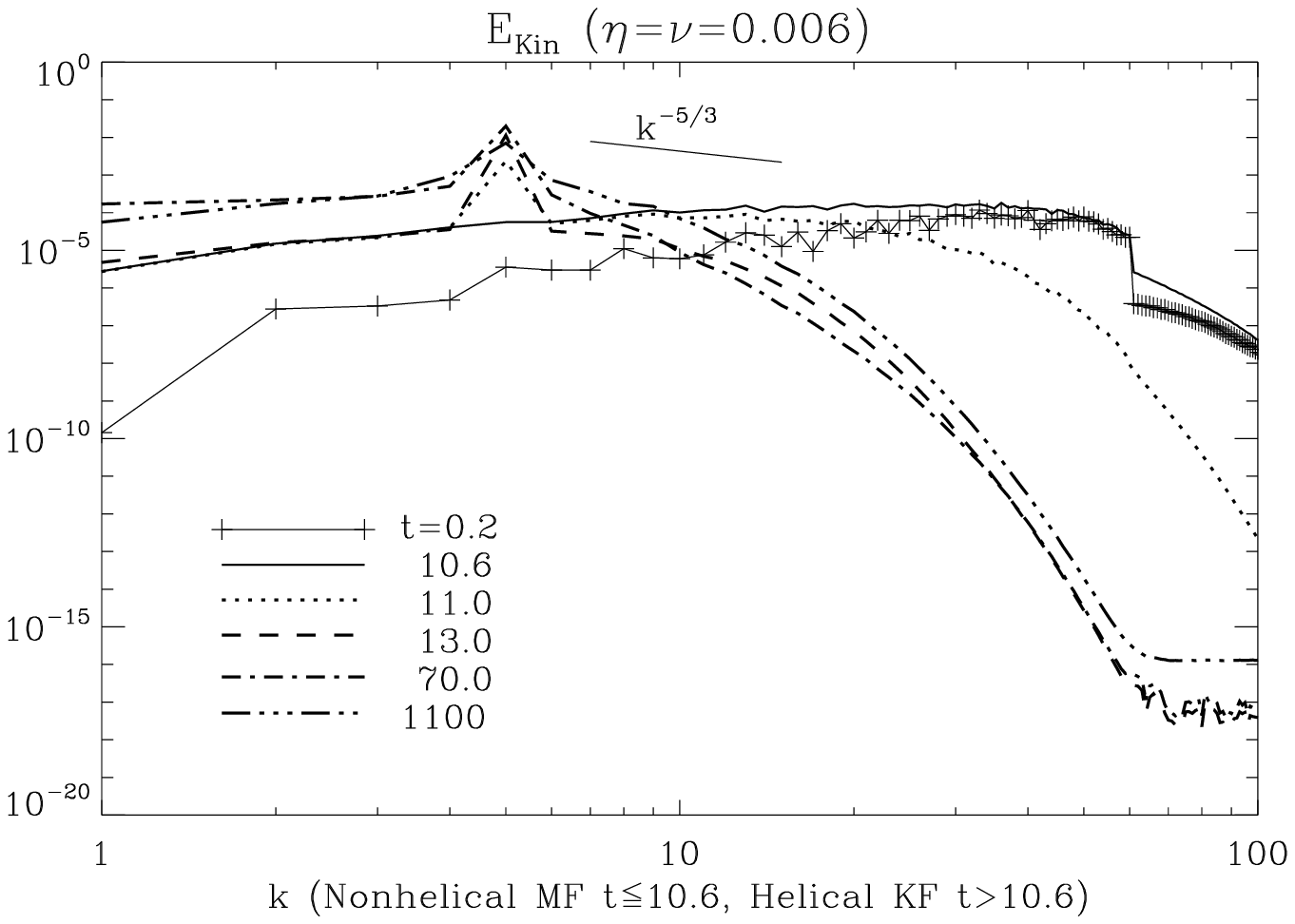}
     \label{4}
   }\,
   \subfigure[$E_{mag}$ ($NHMF\rightarrow HKF$)]{
     \includegraphics[width=8cm]{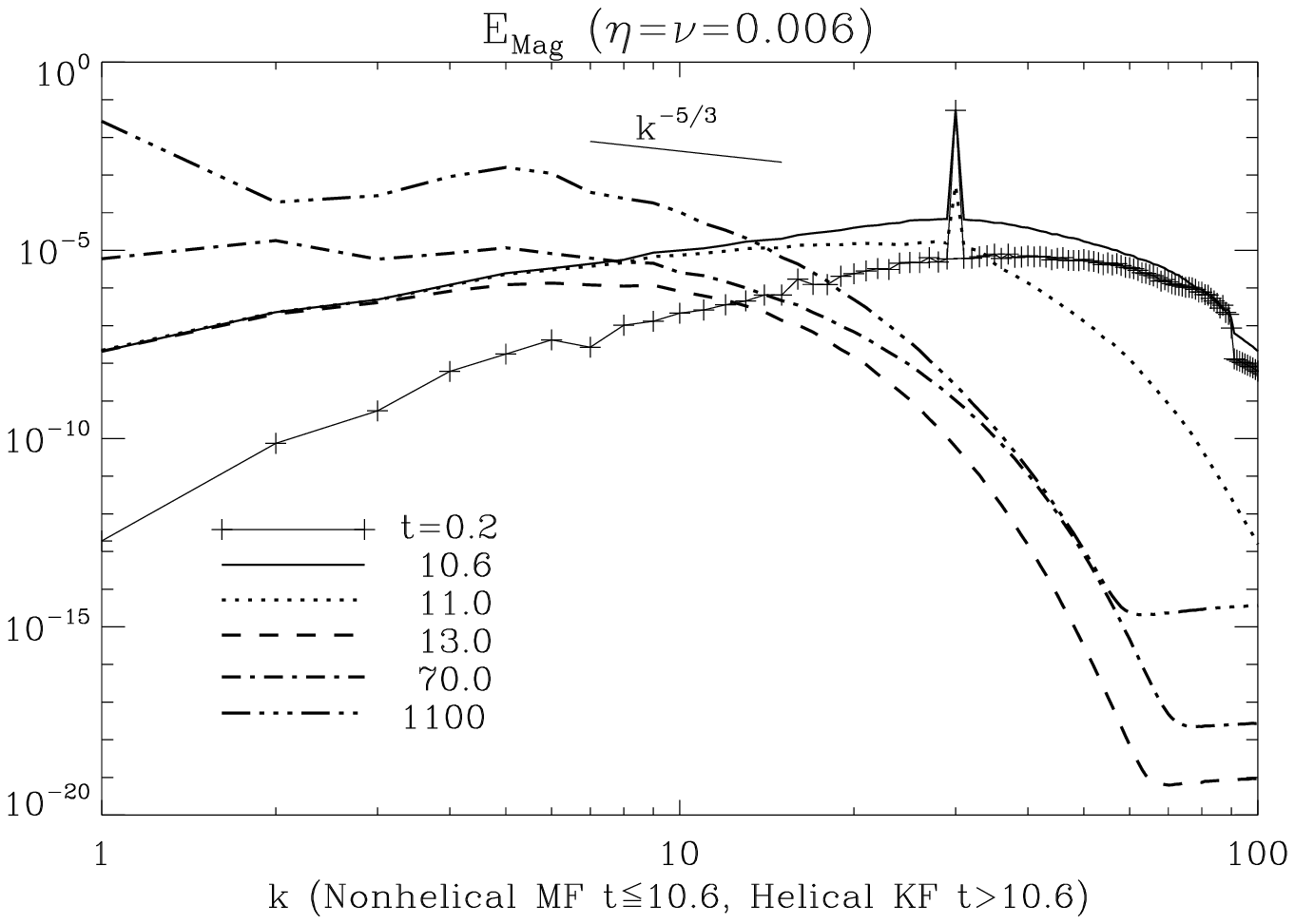}
     \label{5}}\,
   \subfigure[$E_{kin}$ and $E_{mag}$]{
     \includegraphics[width=8cm]{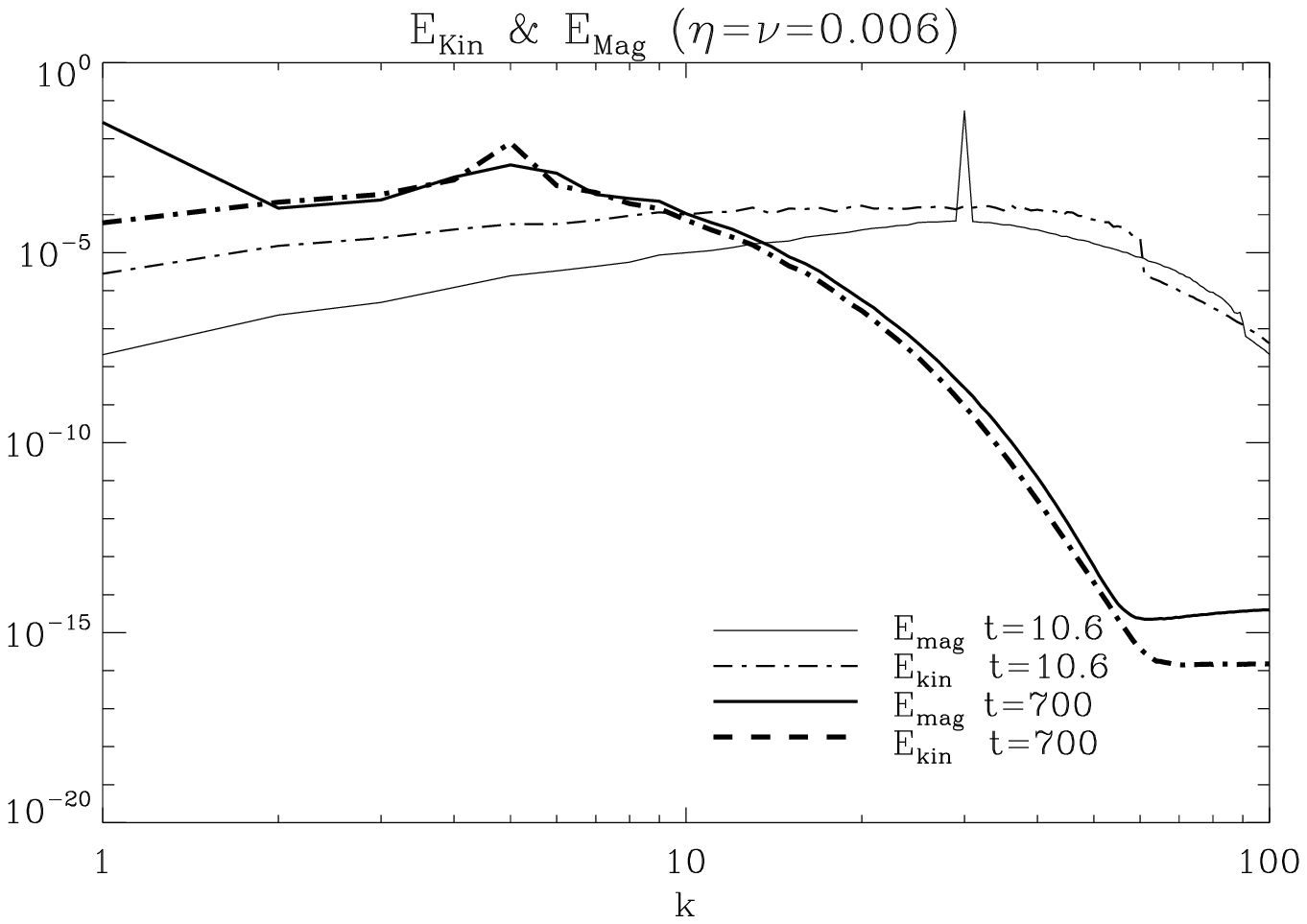}
     \label{6}
   }\,
     }
\caption{$NHMF$($fh_m=0$, $f_0=0.01$, $k_f=30$) finishes at $t=10.6$, and then $HKF$($fh_k=1$, $f_0=0.07$, $k_f=5$) begins. Initially, only tiny $E_M$ is given($E_{kin}$ is $zero$). But $E_{kin}$ grows quickly, catches up with $E_M$ till $t\sim0.2$, and outweighs it. (a) $E_{kin}$ which is transferred from magnetic eddy through Lorentz force migrates backward and forward. (b) The diffusion of energy among magnetic eddies without $\alpha$ effect in $NHMF$ is tiny. Except the forced eddy, the energy in magnetic eddies seems to be mostly from kinetic eddies. After the precursor simulation, the peak of $E_M$(nonhelical) at $k=30$ disappears within a few time steps. (c) Comparison of $E_{kin}$ and $E_M$.}
}
\end{figure}

\begin{figure}
\centering{
{%
   \subfigure[$E_{kin}$ ($HMF\rightarrow HKF$)]{
     \includegraphics[width=8cm]{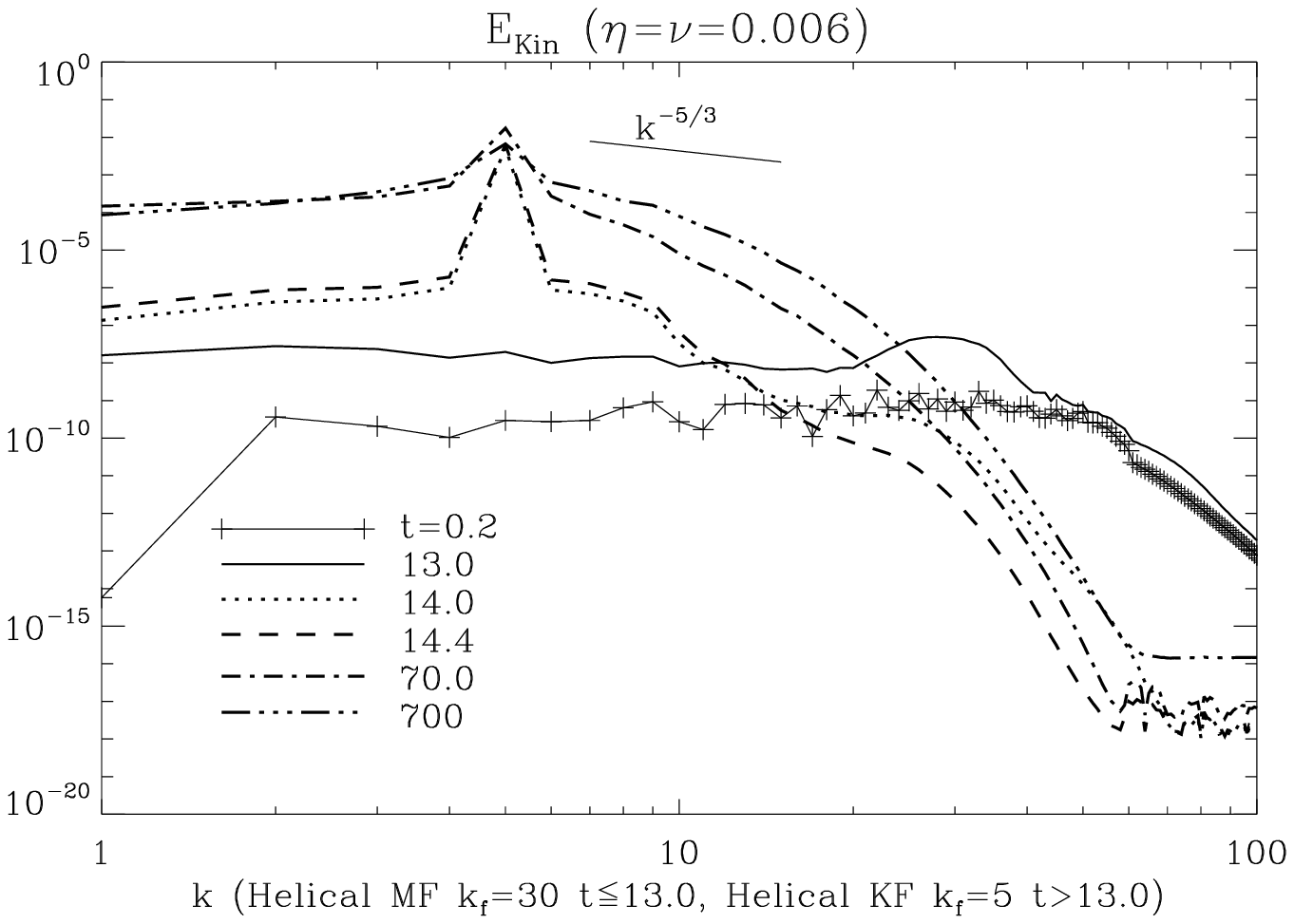}
     \label{7}
   }\,
   \subfigure[$E_{mag}$ ($HMF\rightarrow HKF$)]{
     \includegraphics[width=8cm]{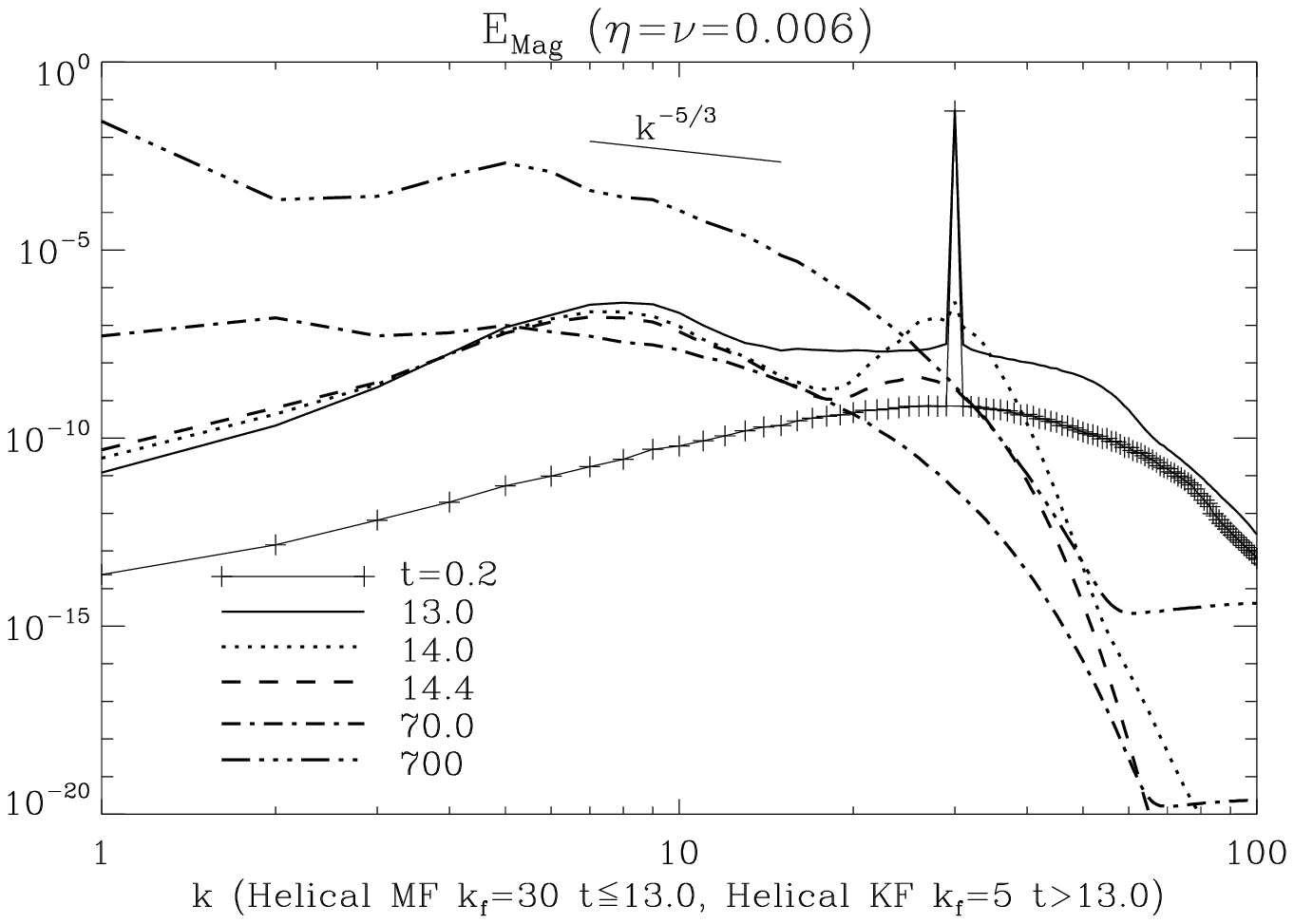}
     \label{8}}\,
   \subfigure[$E_{kin}$ and $E_{mag}$]{
     \includegraphics[width=8cm]{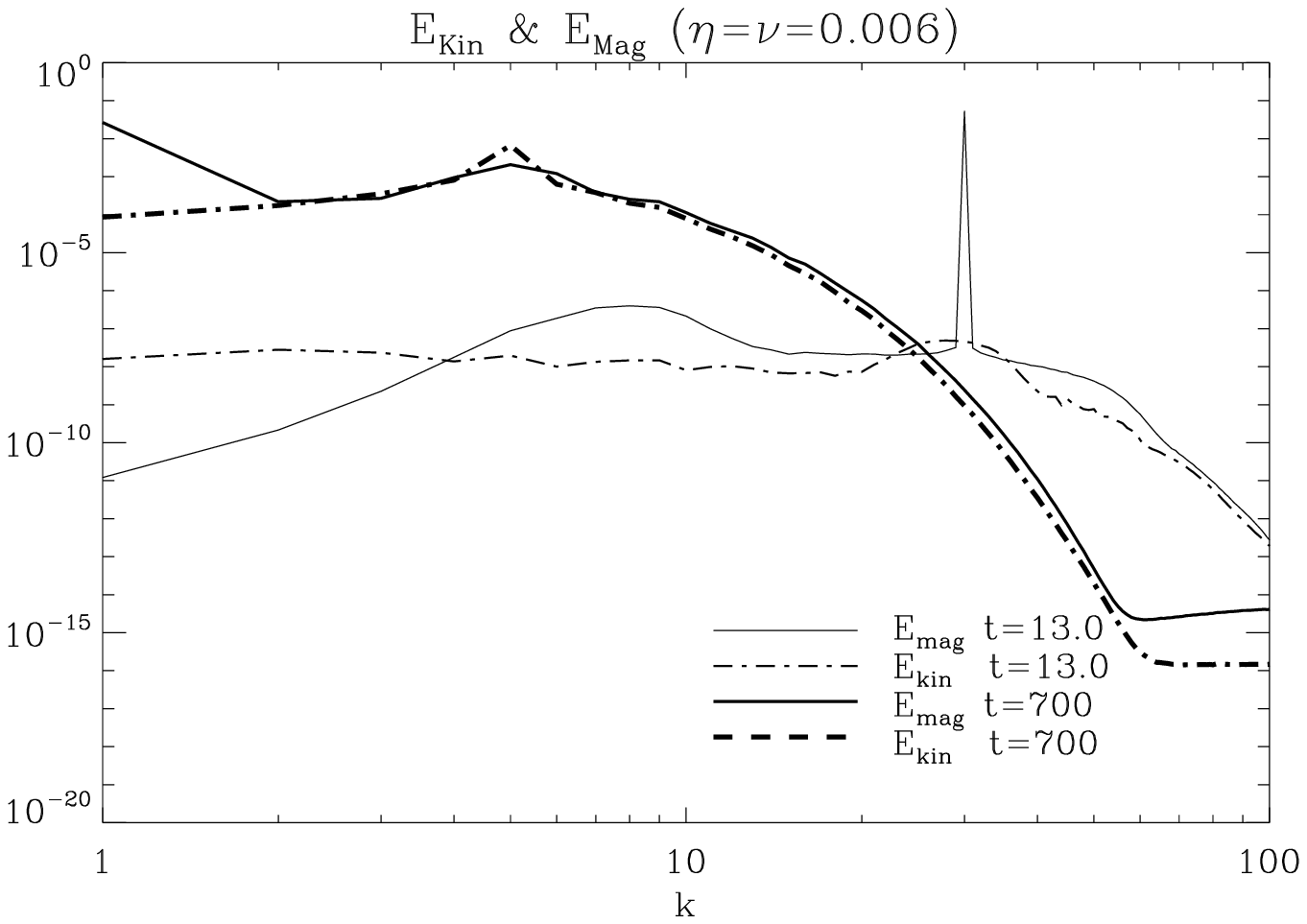}
     \label{9}
   }\,
     }
\caption{$HMF$($fh_m=1$, $f_0=0.01$, $k_f=30$) finishes at $t=13.0$, and then $HKF$($fh_k=1$, $f_0=0.07$, $k_f=5$) begins. (a) $E_{kin}$ of $HMF$ is smaller than that of $NHMF$. (b) $E_{mag}$ of $HMF$ is also smaller than that of $NHMF$. The second small peak around $k=9,\,10$ is the inversely cascaded energy due to $\alpha$ effect. This peak moves backward to be merged into the new forcing peak($k=5$) when $HKF$ begins. The peak of $E_M$ at $k_f=30$ also disappears within a few time steps.}
}
\end{figure}

\begin{figure}
\centering{
{%
   \subfigure[]{
     \includegraphics[width=8cm]{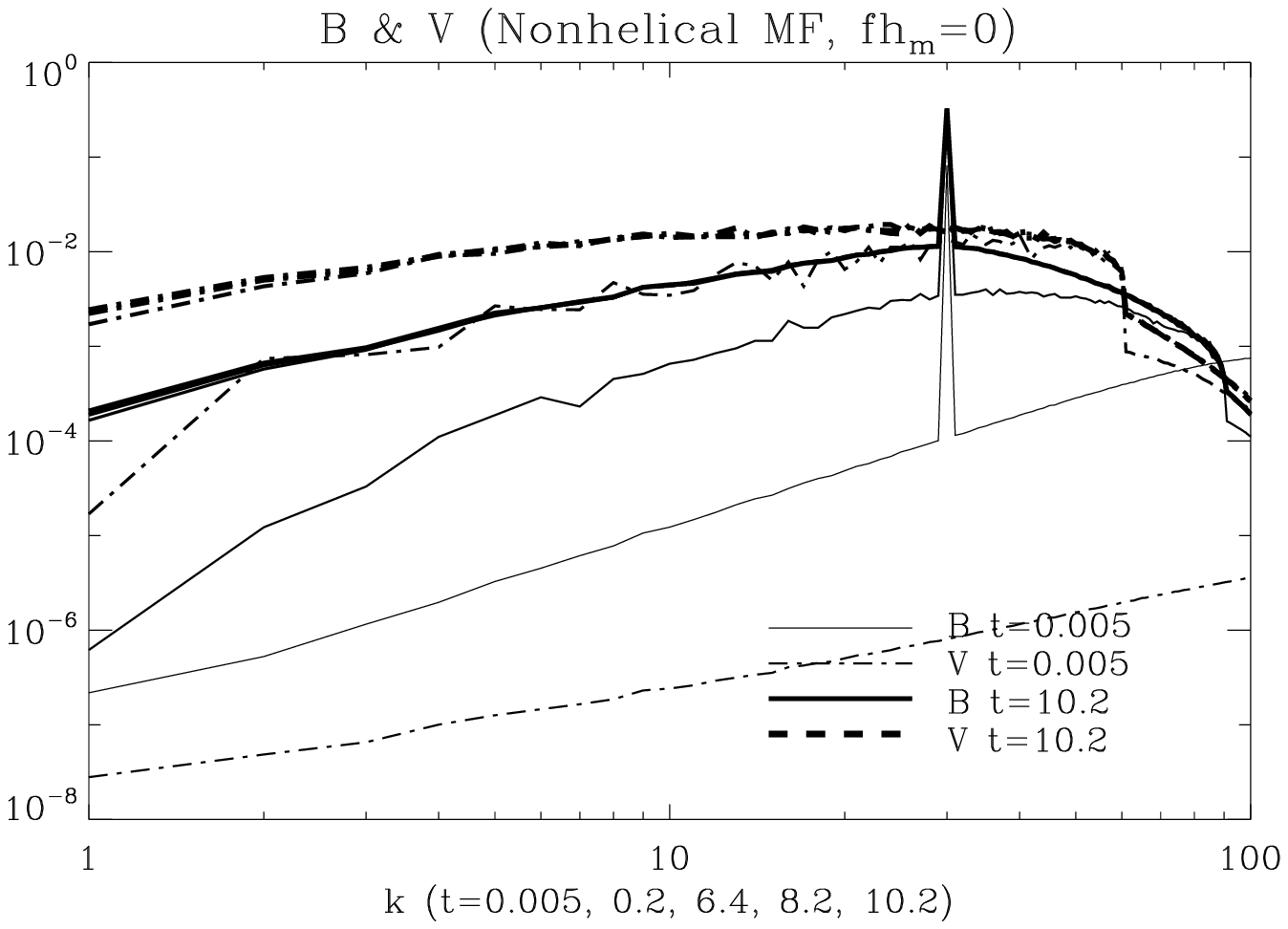}
     \label{10}
   }\,
   \subfigure[]{
     \includegraphics[width=8cm]{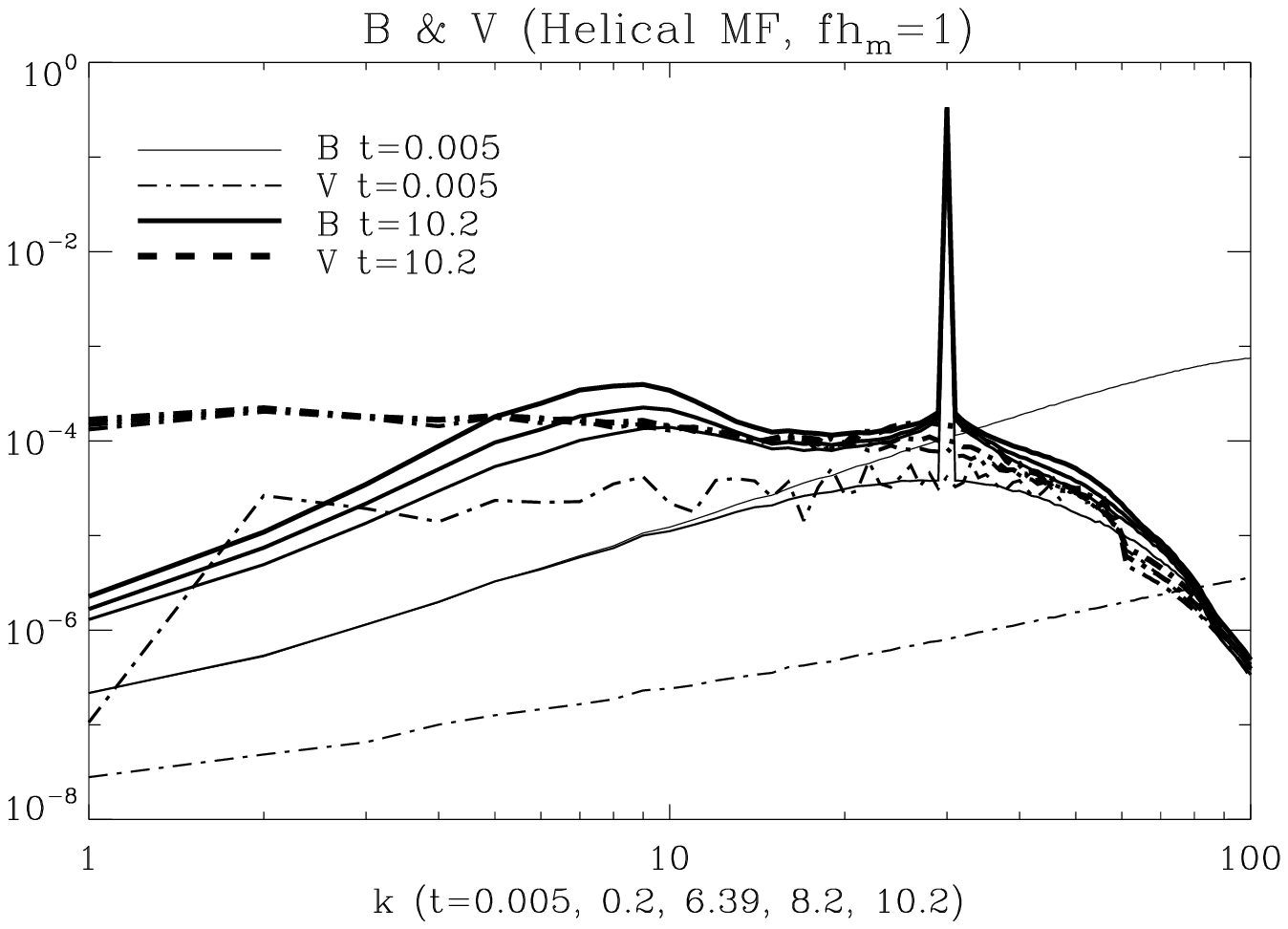}
     \label{11}
   }\,
   \subfigure[]{
     \includegraphics[width=8cm]{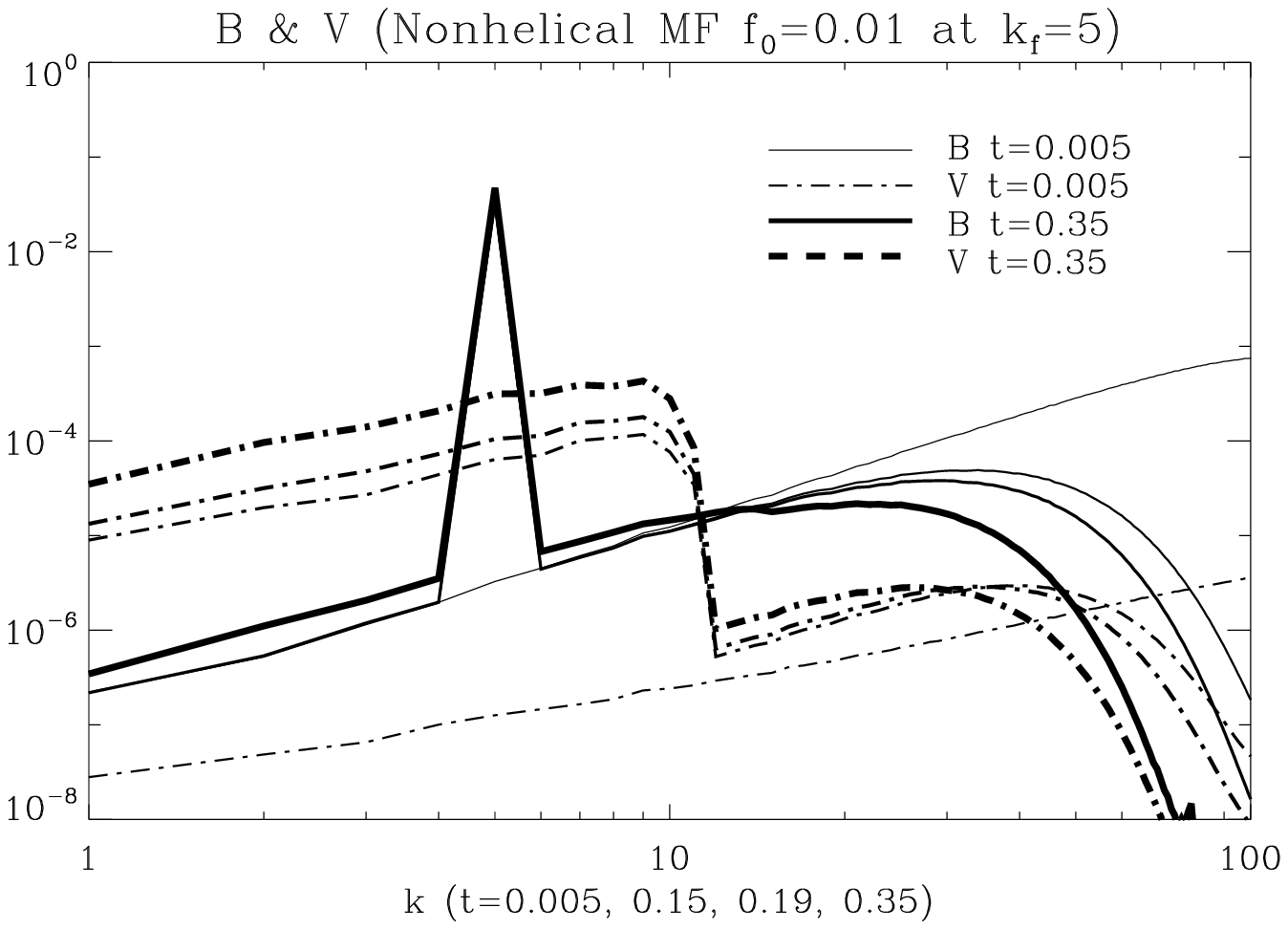}
     \label{12}
   }
     }
\caption{(a), (b) are the preliminary simulation before $HKF$. (c) This plot is the same as that of (a), but the forced eddy is $k_f=$5, closer to the large scale. It shows basic profile of $E_{kin}$ does not so much depend on the position of forced eddy. Also linear and uniform kinetic energy distribution implies the energy transfer is more local and contiguous rather than nonlocal.}
}
\end{figure}

Fig.1 includes the early time profiles of large scale $|H_M|$(solid line) and $E_M$(dotted line). In Fig.\ref{1}, top line group includes $|H_M|$ and $E_M$ for the case of $NHMF\rightarrow HKF$: after Non Helical Magnetic Forcing(as a precursor simulation, $k_f$=30 t$\leq$$10.6$) Helical Kinetic Forcing($k_f$=5, $t$$>$$10.6$) was done over this preliminary simulation. The middle lines are $|H_M|$ and $E_M$ for $HMF\rightarrow HKF$: Helical Magnetic Forcing($k$=30, $t$$\leq$$13.0$) $HKF$($k_f$=5, $t$$>$$13.0$). The lowest lines are $|H_M|$ and $E_M$ for $HKF$ system at $k_f$=5 as a reference simulation. The spectra show $NHMF\rightarrow HKF$ is the most efficient in the growth of large scale $H_M$ and $E_M$ in the early time regime. $HKF$ of which energy transfer chiefly depends on $\alpha$ effect appears to be the least efficient in energy transfer. In case of $HMF\rightarrow HKF$, the efficiency is between `$NHMF\rightarrow HKF$' and `$HKF$'. And during $HMF$, the features of $NHMF$ and $HKF$ due to the helical and nonhelical field are observed.
\\

\noindent The spectrum of magnetic energy $E_M$ is always positive, but the sign of $H_M$ is influenced by the external driving function `$f$' and forcing method. For example, in case of $HMF$, $H_M$ and `$f$' have the same sign(middle lines, $t\leq$$13.0$), but in case of $HKF$, $H_M$ has opposite sign of `$f$'. The cusp in this group($t\sim13$) is the rapid change of $H_M$ from positive($HMF$, $fh_m$=$\langle k\,{\bf a}$$\cdot$${\bf b}
\rangle/\langle {\bf b}^2\rangle$=$+1$, full helical) to negative($HKF$, $fh_k$=$\langle {\bf v}$$\cdot$${\bf \omega}
\rangle/\langle k\,{\bf v}^2\rangle$=+$1$, full helical). The reasons of reversed or equal sign of large scale magnetic helicity in $HKF$ and $HMF$ were partially explained in \cite{2012MNRAS.419..913P}, \cite{2012MNRAS.423.2120P}. In contrast, the magnitude of $|H_M|$ tends to be continuous, which implies the relation between $H_M$ and $E_M$. This will be dicussed again. On the contrary, the direction and magnitude of $H_M$ in $NHMF$ are irregular because of the fluctuating `$f$'. During $NHMF$(t$\leq$$10.6$) the actual sign of $H_M$ is negative. However, if $NHMF$ keeps going on, $H_M$ will change the sign slowly and irregularly.
\\

\noindent Fig.\ref{2} includes the linearly scaled plots in Fig.\ref{1}. All simulations started with the same seed field. However, as the plots show, the system driven by $NHMF$ has the largest and fastest growing $E_{M}$ and $H_{M}$ in the early time regime. In addition, during $HKF$ after the preliminary $NHMF$, $E_{M}$ and $H_{M}$ still grow fastest, and maintain the largest values until they get close to the saturation. This shows the effects of $IC$s and forcing method clearly.\\

\noindent Fig.\ref{3} is to compare directly the onset and saturation of $E_M$ and $|H_M|$. The group of three lines in the left part has the original data plots, and the lines in the right group are their shifted plots for comparison. The onset point of $NHMF\rightarrow HKF$ is the earliest, followed by $HMF\rightarrow HKF$, and then $HKF$. The order of onset position is closely related to the different amount of magnetic energy and helicity generated during the preliminary simulation. However, as Table.1 shows, $H_{M, L}$($L$: large scale), $E_{M, L}$, $E_{kin, L}$, and $E_{kin, s}$($s$: small scale) in each case are of similar values at their onset positions in spite of the different $IC$s. They are sort of critical values for the onset, and the time to reach this critical point is inversely proportional to the magnitude of $IC$s. Since the forcing method after the preliminary simulation is the same ($HKF$), $IC$s are determinants of the onset time. However, the saturation of turbulence dynamo is independent of $IC$s. Rather the saturation is decided by the external forcing `$f$' and intrinsic properties of system like viscosity $\nu$($\sim$$Re$$^{-1}$, Reynolds number $Re=V_{rms}L/\nu$. L: characteristic linear dimension) or magnetic diffusivity $\eta$($\sim$$Re_M$$^{-1}$, Magnetic Reynolds number $Re_M=V_{rms}L/\eta$). All three simulations have the same saturated $Re_M$$\sim$30. Those features of critical values and saturation imply there is no long lasting memory effect in turbulence. This validates Markovianization in MHD equations for closure.
\\

\noindent Fig.\ref{4}$\sim$\ref{6} are the spectra of $E_{kin}$ and $E_M$ of $NHMF\rightarrow HKF$. During the preliminary $NHMF$ at $k_f=30$, most $E_M$ is localized at the forced eddy (Fig.\ref{5}). But $E_{kin}$ which is larger than $E_M$ in most range spreads over from large to small scale(Fig.\ref{4}, \ref{6}). This relatively linear profile of $E_{kin}$ spectrum indicates the pressure that makes the system homogeneous transporting the energy forward and backward is dominant in the very early time regime. On the other hand, these figures, especially Fig.\ref{6}, imply the relation between $E_{kin}$ and $E_M$. Initially only tiny seed $E_M$ was given to the system. However, once $NHMF$ started, $E_{kin}$ caught up with $E_M$ by $t\sim0.2$ and outweighed it. Soon, larger $E_{kin}$ gets to induce $E_M$ which is the source of $H_M$. That $E_{kin}$ is one of the sources of $E_M$ is coincident with the result of EDQNM approximation(\cite{1976JFM....77..321P}). On the other hand, the backward transfer of kinetic energy seems to contradict the accepted theorem that inverse cascade in three dimensional magnetohydrodynamic turbulence is not possible. However, when the energy or vorticity is not conserved, $E_{kin}$ can be inversely cascaded. We will come back to this problem later.
\\

\noindent Fig.\ref{7}$\sim$\ref{9} are $E_M$ and $E_{kin}$ for $HMF\rightarrow HKF$. $HMF$ shows two kinds of energy transports: nonlocal transport of $E_M$ and local transport of $E_{kin}$. The former is caused by helical field($\alpha$ effect), and the latter is caused by the pressure. In Fig.\ref{7}, backward migration of $E_{kin}$ indicates the role of pressure. And the increase of $E_M$ in large scale implies the direct energy transfer from kinetic eddy(Fig.\ref{8}, \ref{9}). Furthermore, the helical driving source generates current helicity $\langle \bf j\cdot \bf b \rangle$(=$\langle k^2\bf a\cdot \bf b \rangle$) and kinetic helicity $\langle \bf v\cdot \bf \omega \rangle$, which forms $\alpha$ effect in the system. This $\alpha$ effect generates the secondary peak of $H_M$ around $k\sim10$(Fig.\ref{8}). It keeps moving backward and merges into the main peak when $HKF$ begins at $k_f=5$.\\

\noindent The difference between Fig.\ref{10}($NHMF$) and Fig.\ref{11}($HMF$) is just magnetic helicity ratio($fh_m=0$, $fh_m=1$). Fig.\ref{10} and Fig.\ref{12}($NHMF$) have the same helicity ratio $fh_m=0$ but different forced eddies($k_f=30$, $k_f=5$). These clearly show kinetic energy migration and the basic profile of field evolution do not depend on the position of forced eddy so much.
\\

\begin{figure}
\centering{
  {
   \subfigure[]{
     \includegraphics[width=8cm]{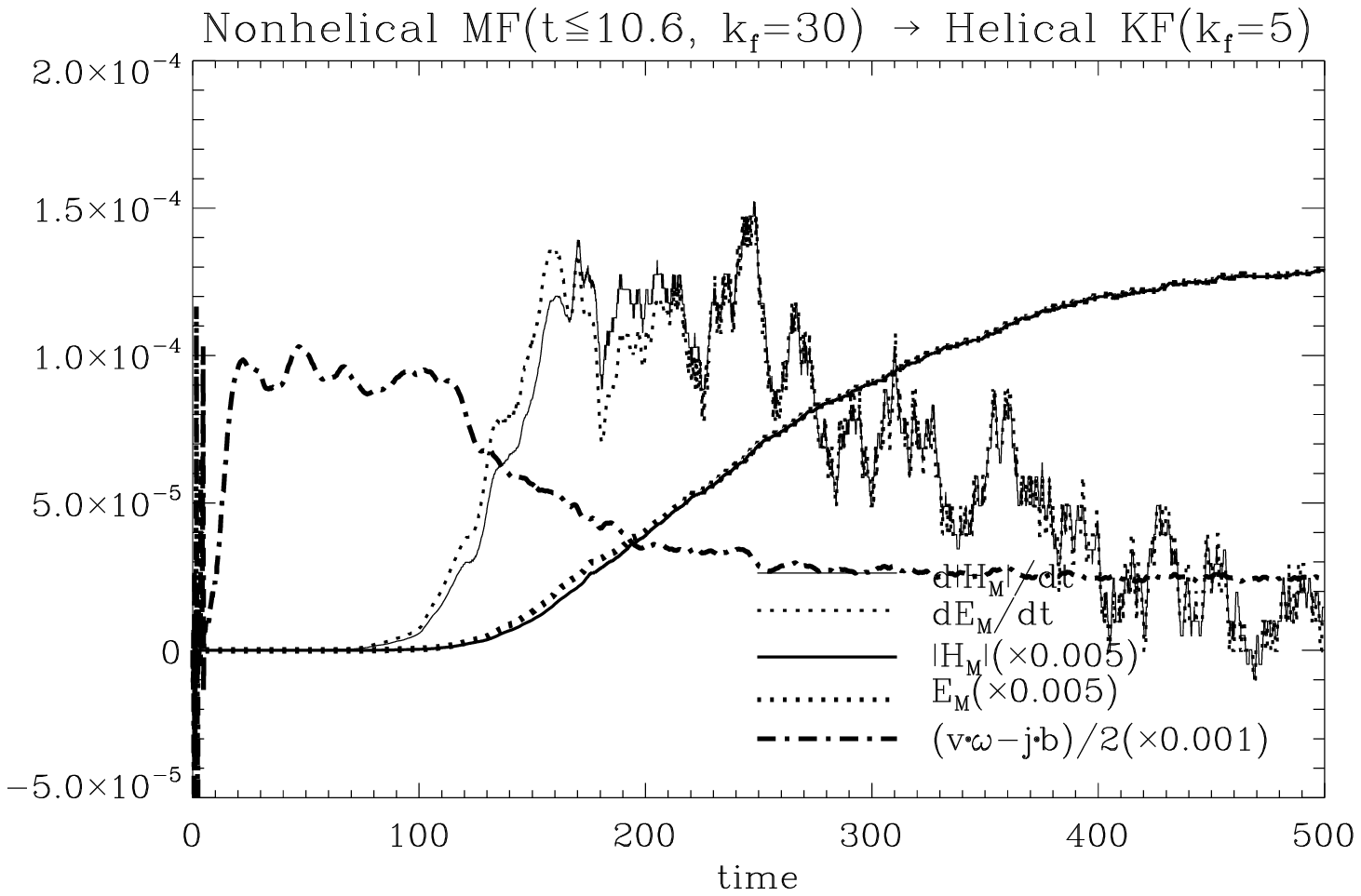}
     \label{13}
   }\,
   \subfigure[]{
     \includegraphics[width=8cm]{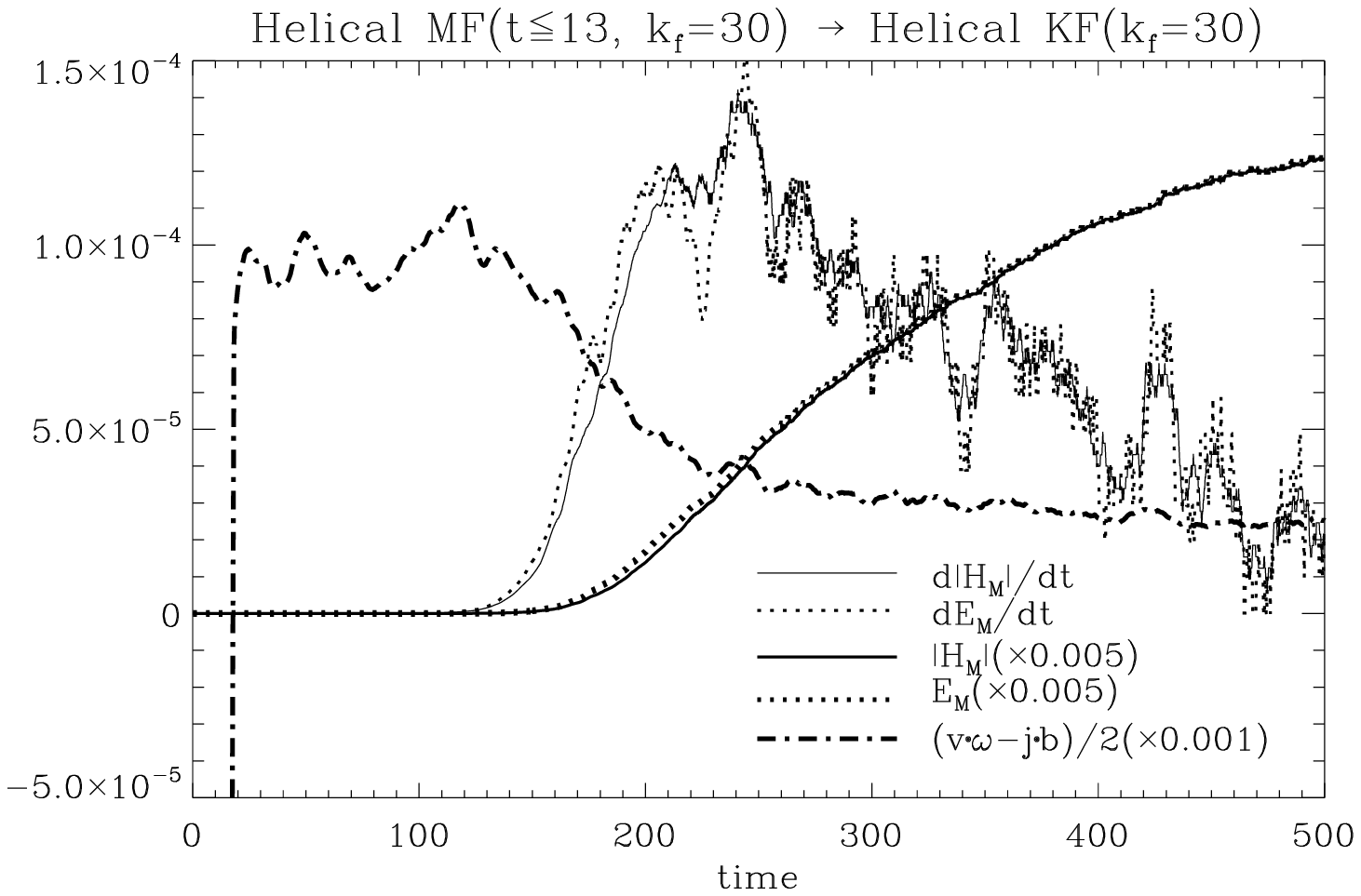}
     \label{14}}\,
   \subfigure[]{
     \includegraphics[width=8cm]{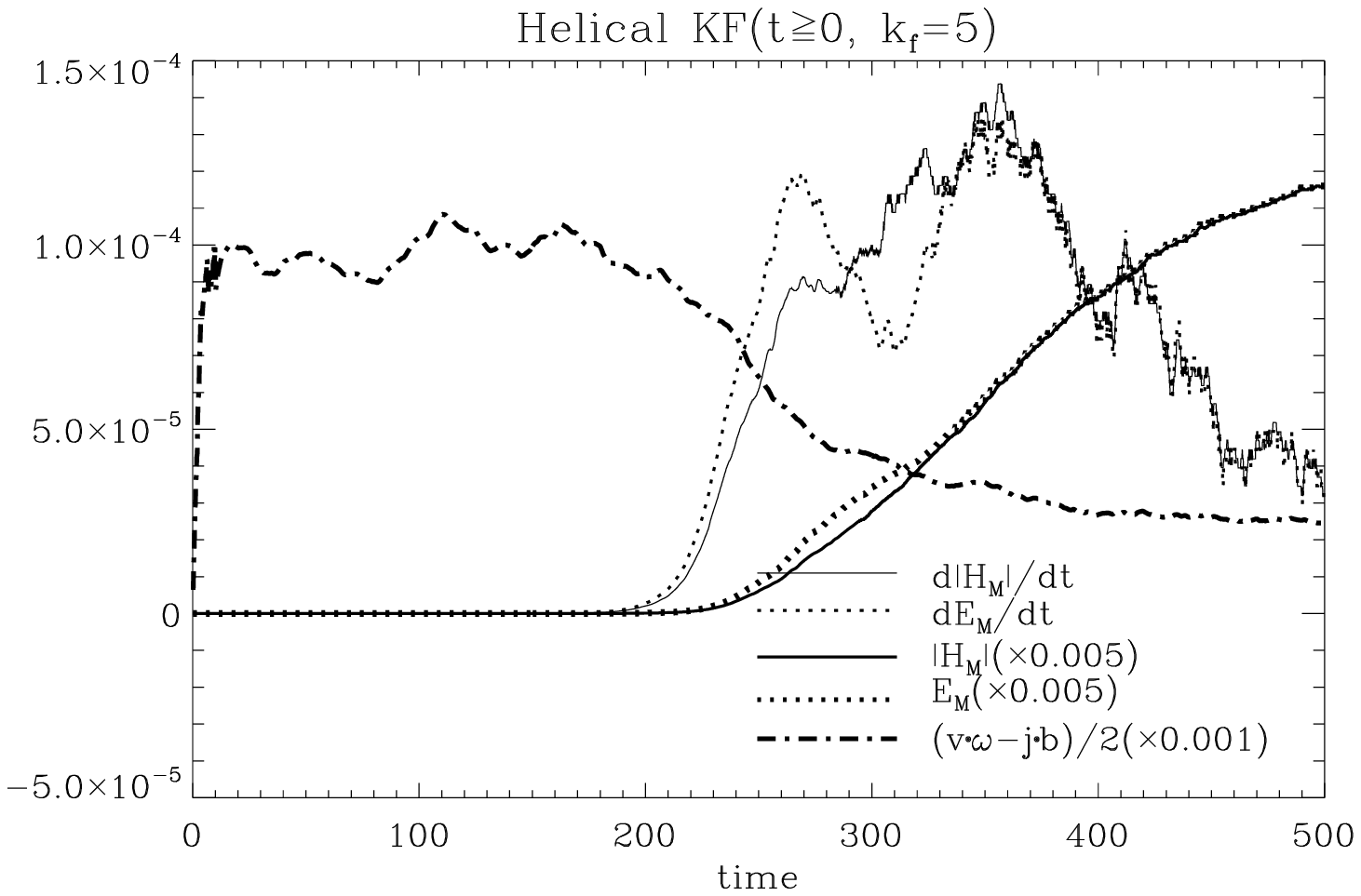}
     \label{15}
   }\,
  }
\caption{During $HKF$, ($\langle\bf v$$\cdot$$\omega\rangle$-$\langle\bf j$$\cdot$ $\bf b\rangle$)/2 in each case drops at different time position. It depends on the energy and helicity($IC$s) from the preliminary simulation. Also due to the different eddy turnover time between large and small scale, there is a phase difference in the profile of growth rate, $E_M$($H_M$), and $\alpha$ related term. For the growth ratio, usually logarithmic growth ratio is used: $d\,log |H_M|/dt=-\alpha^R E_M/|H_M|-2k^2\nu_v\,(k=1)$, but linear growth rate was used for the mathematical convenience and visibility. All quantities but $E_M$ and $H_M$ are the averages of 50$\sim$100 nearby values.}
}
\end{figure}

\noindent Fig.\ref{13}$\sim$\ref{15} include the profiles of $dH_M/dt$, $dE_M/dt$, $E_M$$(\times 0.005)$, $H_M$$(\times 0.005)$, and ($\langle\bf v$$\cdot$$\omega\rangle$-$\langle\bf j$$\cdot$$\bf b\rangle$)/2($\times 0.001$). These plots are helpful to infer their relative roles and relations in large scale dynamo(Eq.(\ref{Hm Equation}), (\ref{Em Equation})). The profile of ($\langle\bf v$$\cdot$$\omega\rangle$-$\langle\bf j$$\cdot$$\bf b\rangle$ $\sim$$-\alpha$) shows the effects of `$f$' and $IC$s clearly. With the smallest $IC$s in $HKF$(Table.1), its duration time($0<t<\sim200$, Fig.\ref{15}) of constant $\alpha$ coefficient is longer than that of other cases. In contrast the simulation of `$NHMF\rightarrow HKF$' has the shortest duration time of constant $\alpha$ coefficient($\sim20<t<\sim110$, Fig.\ref{13}). As an another feature, $dE_M/dt$ is not always larger than $d|H_M|/dt$.
\\

\begin{figure*}
\centering{
\mbox{%
   \subfigure[]{
     \includegraphics[width=8cm]{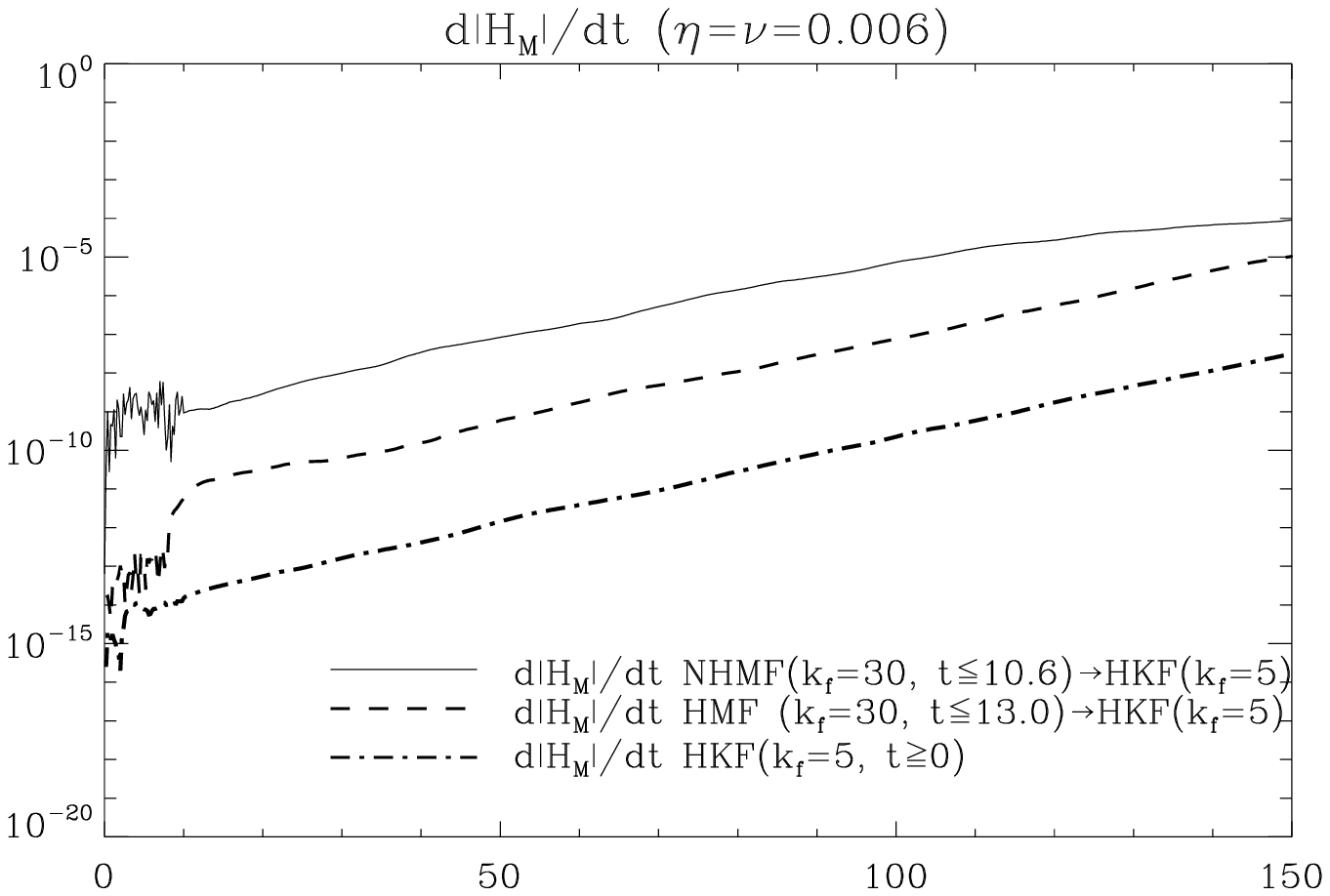}
     \label{16}
   }\,
   \subfigure[]{
     \includegraphics[width=8cm]{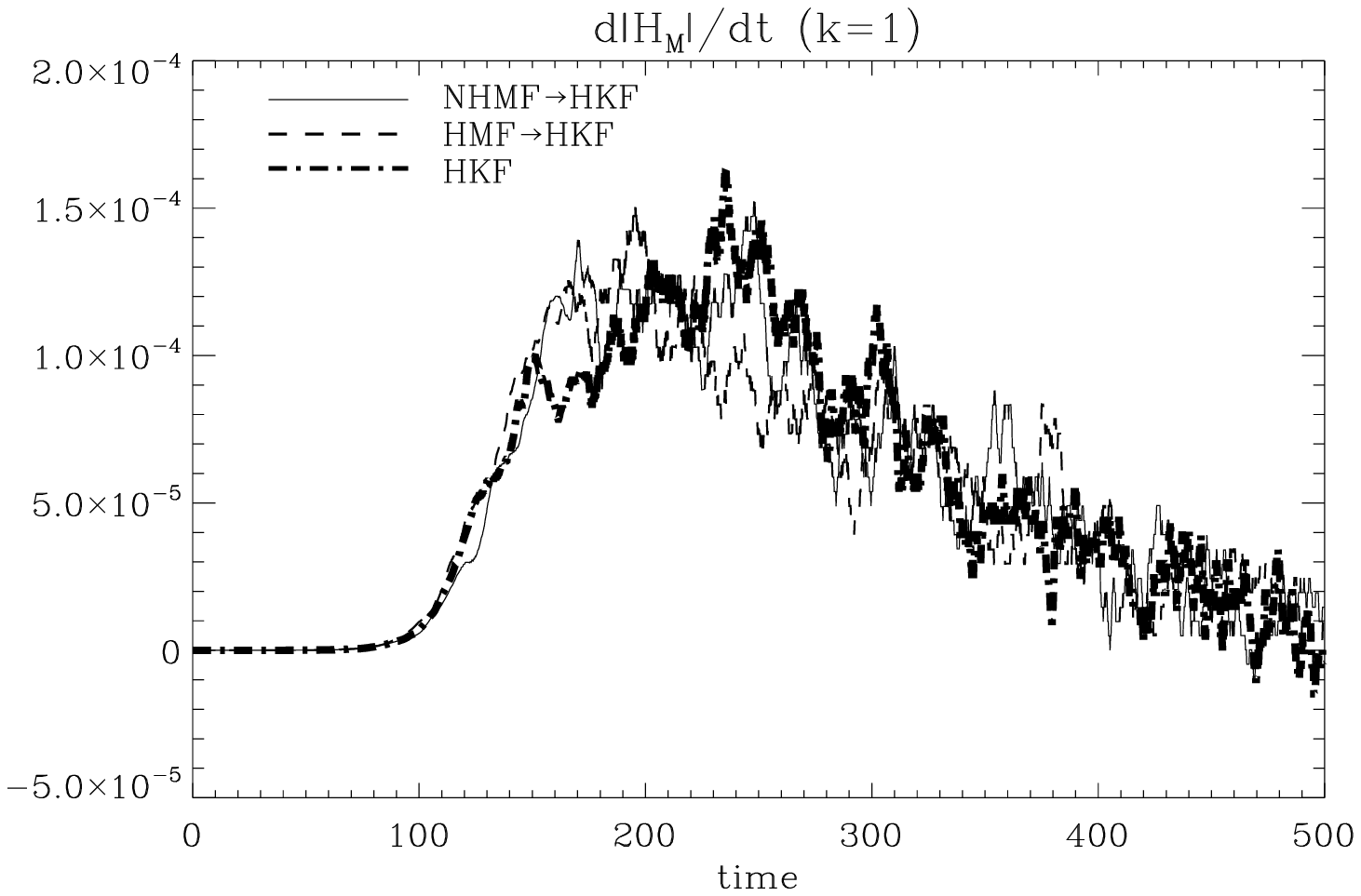}
     \label{17}}
     }
\caption{(a) Growth ratio proportionally depends on the $IC$s. The area between the line and time axis is $H_M$. (b) The profile of evolving growth ratio of $HKF$ is slightly different from that of others. It seems to be caused by the turbulent effect. Each growth ratio is the average of 50 nearby points.}
}
\end{figure*}

\noindent Fig.\ref{16} shows the profiles of $d|H_M|/dt$ in the early time regime. And $d|H_M|/dt$ in Fig.\ref{17} are shifted plots for the comparison. All $d|H_M|/dt$ converge to zero; but, the profile of $HKF$ follows different paths. $d|H_M|/dt$ of $HKF$ is smaller than that of the other cases until it reaches the onset position. $HKF$ starts with the smallest $IC$s, but all quantities except $E_{M, s}$ become the same as those of other cases by the onset position. When the field is about to arise, $E_{M, s}$ of $HKF$($\sim 4.3\times 10^{-5}$) is smaller than that of other cases($\sim7\times 10^{-5}$). In theory, this term is discarded because of the seemingly little influence on the evolution of $E_M$ or $|H_M|$ (\cite{1976JFM....77..321P}, \cite{2002ApJ...572..685F}, \cite{2002PhRvL..89z5007B}). However, $E_{M,s}$ is closely related to the conservation of magnetic helicity in the system and constraining velocity field. We will discuss about this again.

\section{Theoretical model}
There is no theoretical method that can completely explain the influence of $IC$s like $E_{kin}$, $E_M$, or $H_M$ on MHD dynamo yet. However, some approximation like $EDQNM$(\cite{1976JFM....77..321P}), though limited, can be used. The representations of $H_M$ and $E_M$ of this method are quite similar to those of two scale mean field method(\cite{2004PhPl...11.3264B}, \cite{2002ApJ...572..685F}). The equations are composed of Alfv$\acute{e}$n effect term by the larger eddies, $\alpha$ effect term by the smaller eddies, and dissipation term. These approximate equations assume the field is composed of helical and nonhelical part. If helical component in the field is $zero$ or ignorably small($NHMF$ or $NHKF$), these equations are not valid. The system is divided into large($k=1$) and small scale($k=2\sim k_{max}$), and this small scale can be subdivided into the forcing($k=2\sim 6$) and smaller scale($k=7\sim k_{max}$).\\
\begin{eqnarray}
\frac{\partial H_M}{\partial t}&\cong&\big(\overbrace{(\Gamma/k)(H_v-k^2H_M)+\widetilde{\Gamma}E_v}^{Alfv\acute{e}n\,\, effect}\big)+
\alpha^R E_M-2\nu_v k^2 H_M\nonumber\\
&\cong&\alpha^R E_M-2\nu_v k^2 H_M,
\label{Hm Equation}
\end{eqnarray}
\begin{eqnarray}
\frac{\partial E_M}{\partial t}&\cong&\big(\overbrace{k\Gamma(E_v-E_M)+\widetilde{\Gamma}H_v}^{Alfv\acute{e}n\,\, effect}\big)+\alpha^R k^2 H_M-2\nu_v k^2 E_M\nonumber\\
&\cong&\alpha^R k^2 H_M-2\nu_v k^2 E_M \quad (E_{kin}\equiv E_v).
\label{Em Equation}
\end{eqnarray}
\\
\noindent The coefficients are,
\begin{eqnarray}
\alpha^R&=&-\frac{4}{3}\big[\int^{\infty}_{k/a}\theta_{kpq}(t)\big(H_v(q)-q^2H_M(q)\big)dq\big], \quad (a<1)\nonumber\\
\nu_v&=&\frac{2}{3}\int^{\infty}_{k/a}\theta_{kpq}(t)E_v(q) dq,\,\,\,\, \theta_{kpq}(t)=\frac{1-\exp(-\mu_{kpq}t)}{\mu_{kpq}},\nonumber\\
\mu_{k}&=&\mathrm{C_s}\big[\int^k_0q^2(E_v(q)+E_M(q))dq\big]^{1/2}\nonumber\\
&&+(1/\sqrt{3})k\big[2\int^k_0 E_M(q)dq\big]^{1/2}+(\nu+\eta)k^2,\nonumber\\
\Gamma&=&\frac{4}{3}k\int^{ak}_{0}\theta_{kpq}(t)E_M(q)dq,\nonumber\\
\widetilde{\Gamma}&=&\frac{4}{3}\int^{ak}_{0}\theta_{kpq}(t)q^2H_M(q) dq.
\label{kinetic nu theta mu definition}
\end{eqnarray}

\noindent $H_v$ is kinetic helicity(=1/2$\langle\bfv$$\cdot$$\omega \rangle$), $H_M$ is magnetic helicity(=$1/2\abl$, please note the coefficient), and $\nu_v$ is kinetic eddy diffusivity. $\alpha^R$ that transfers $H_M$ and $E_M$ to larger scale is composed of the residual helicity($q^2H_M(q)-H_v(q)$) and triad relaxation time $\theta_{kpq}$. $\theta_{kpq}$ is the function of eddy damping rate(see appendix) $\mu_{kpq}$$(=\mu_{k}+\mu_{p}+\mu_{q})$, and connects smaller scale eddies and larger scale eddies.\\

\noindent The influence of Alfv$\acute{e}$n terms($k=0$) on the large scale($k=1$) is physically meaningless. Ignoring Alfv$\acute{e}$n terms($k=0$), we find those coupled equations have two normal coordinates: `$E_M+H_M$' and `$E_M-H_M$'.
\begin{eqnarray}
\frac{\partial (E_M+H_M)}{\partial t}&=&(\alpha^R-2\nu_v)(E_M+H_M),
\label{E+H Equation}
\end{eqnarray}
\begin{eqnarray}
\frac{\partial (E_M-H_M)}{\partial t}&=&-(\alpha^R+2\nu_v)(E_M-H_M).
\label{E-H Equation}
\end{eqnarray}
Assuming $\int^t_0(\alpha^R-2\nu_v)dt\equiv(\alpha-2\overline{\nu})t$, the solution is
\begin{eqnarray}
H_M(t)&=&\frac{1}{2}\big[H_{M0}\big(e^{(\alpha-2\overline{\nu})t}+e^{-(\alpha+2\overline{\nu})t}\big)\nonumber\\
&&+E_{M0}\big(e^{(\alpha-2\overline{\nu})t}-e^{-(\alpha+2\overline{\nu})t}\big)\big],
\label{EDQNM solution H}
\end{eqnarray}
\begin{eqnarray}
E_M(t)&=&\frac{1}{2}\big[E_{M0}\big(e^{(\alpha-2\overline{\nu})t}+e^{-(\alpha+2\overline{\nu})t}\big)\nonumber\\
&&+H_{M0}\big(e^{(\alpha-2\overline{\nu})t}-e^{-(\alpha+2\overline{\nu})t}\big)\big].
\label{EDQNM solution E}
\end{eqnarray}
These solutions show how $H_M$ and $E_M$ are generated. For example, in case of $H_M(t)$ both $H_{M0}$ and $E_{M0}$ are sources of $H_M(t)$, but $E_{M0}$ produces $H_M(t)$ like an auxiliary source($\sim$Sinh). In the early time regime the effect of $E_{M0}$ on $H_M(t)$ is tiny, but finally becomes on a level with $H_{M0}$. $H_{M, sat}$ converges to $E_{M, sat}$(Fig.\ref{3}) as $t$$\rightarrow$$\infty$. This solution shows that as long as $\alpha$ is larger than dissipation 2$\bar{\nu}$, large scale magnetic field eventually becomes fully helical by $\alpha$ effect.(\cite{1991PhFlB...3.1848S}, \cite{2008matu.book.....B})\\

\begin{figure*}
\centering{
\mbox{%
   \subfigure[]{
     \includegraphics[width=8cm]{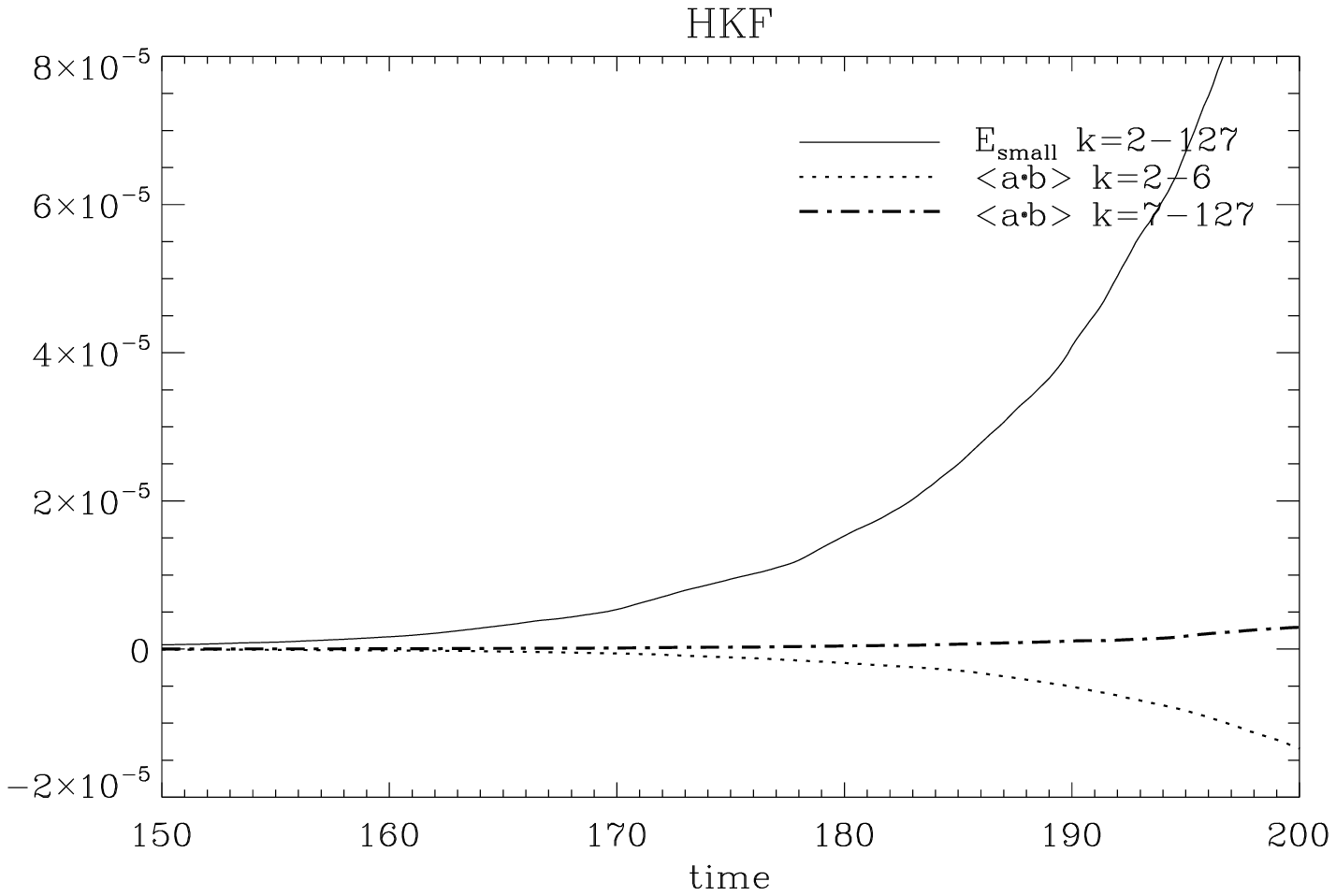}
     \label{18}
   }\,
   \subfigure[]{
     \includegraphics[width=8cm]{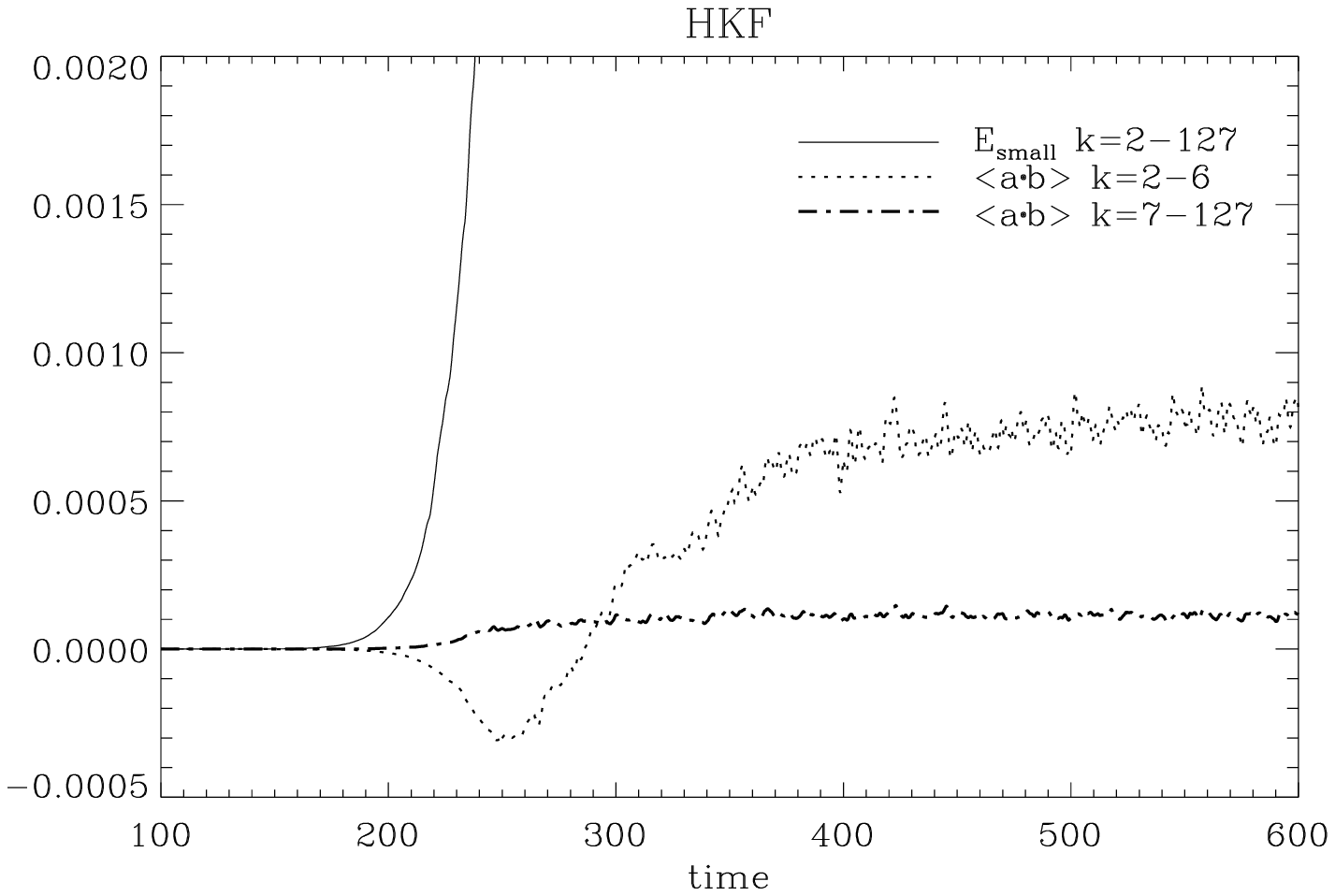}
     \label{19}}
     }
\caption{The direction of magnetic helicity is decided according to the conservation of total magnetic helicity in the system. The minimum (t$\sim$250) and turning point(t$\sim$280) of magnetic helicity in (b) can be compared with the change in growth of large scale $H_M$ in Fig.\ref{15}.}
}
\end{figure*}

\noindent On the other hand, one of our interests is how long the effects of initial values last in turbulence. In \cite{2004ApJ...603..569M}, the influence of imposed large scale magnetic energy on the system was tested(nonhelical kinetic forcing). The strong magnetic field in the large scale was expected to suppress the formation of small scale fields. However, the effect of imposed magnetic energy disappeared soon, and the system eventually followed the external forcing source. Like an oscillator driven by an external driving source, the effect of $IC$s exist only in the early transient mode. In turbulence smaller eddy loses the information faster than large eddy does.
\\

\noindent To see how long the effect of $IC$s lasts, the formal solution of Eq.(\ref{Hm Equation}) may be useful:
\begin{eqnarray}
H_M(t_n)=e^{-\int^{t_n} 2\nu_v(\tau'')d\tau''}\big[\int^{t_n}_0e^{\int^\tau2\nu_v(\tau')d\tau'}\alpha^R(\tau)E_M(\tau)\,d\tau\nonumber\\
+H_M(0)\big].
\label{Hm formal solution}
\end{eqnarray}
Using the trapezoidal method for the integration part with the assumption of $\int^{t_n} 2\nu_v(\tau)d\tau\equiv 2V(t_n)$ and $t_n\equiv n\Delta t$, we find the approximate solution:
\begin{eqnarray}
H_M(t_n)\sim \big[e^{2V(0)-2V(t_n)}\alpha^R(0)E_M(0)\Delta t+e^{-2V(t_n)}H_M(0)\big]+\nonumber&&\\
+\big[e^{2V(t_1)-2V(t_n)}\alpha^R(t_1)E_M(t_1)+e^{2V(t_2)-2V(t_n)}\alpha^R(t_2)E_M(t_2)+&&\nonumber\\
...+e^{2V(t_{n-1})-2V(t_n)}\alpha^R(t_{n-1})E_M(t_{n-1})\big]\Delta t.\nonumber&&
\label{Approximate solution of formal Hm}
\end{eqnarray}
These show all previous results affect the current magnetic helicity in principle. However the influence decreases exponentially, which is coincident with the simulation results. The decaying speed depends on the several factors: energy, helicity, $\nu$, and $\eta$. Of course the actual $\nu_v$ varies with time. But, since $\nu_v$($\sim V$) changes rather smoothly and saturates to a constant, this inference is qualitatively reasonable.
\\

\renewcommand{\tabcolsep}{1.0mm}
\begin{table*}
\begin{tabular}{|c|c|c|c|c|c|c|c|c|c|}
\hline
&\multicolumn{3}{|c|}{$\mathbf{NHMF\rightarrow HKF}$}&  \multicolumn{3}{|c|}{$\mathbf{HMF\rightarrow HKF}$} & \multicolumn{3}{|c|}{$\mathbf{HKF}$}\\ \hline
& {\footnotesize Init.} & {\footnotesize Onset} & {\footnotesize Sat.} & {\footnotesize Init.} & {\footnotesize Onset}& {\footnotesize Sat.} &{\footnotesize Init.} &{\footnotesize Onset} &{\footnotesize Sat.} \\
& {\footnotesize ($t=10.6$)} & {\footnotesize ($t\sim 70$)} & {\footnotesize ($t\rightarrow \infty$)} &{\footnotesize ($t=13.0$)}  & {\footnotesize ($t\sim 120$)}& {\footnotesize ($t\rightarrow \infty$)} & {\footnotesize ($t=0$)}& {\footnotesize ($t\sim 186$)}&{\footnotesize ($t\rightarrow \infty$)} \\
\hline

$H_{M,L}$ & {\footnotesize $-3.5\times 10^{-9} $} & {\footnotesize $-4.5\times 10^{-6} $} &
{\footnotesize $-2.7\times 10^{-2} $} & {\footnotesize $5.4\times 10^{-12}$}& {\footnotesize $-5.1\times 10^{-6}$} & {\footnotesize $-2.7\times 10^{-2}$}& {\footnotesize $-5.7\times 10^{-15}$}& {\footnotesize $-4.9\times 10^{-6}$}&{\footnotesize $-2.6\times 10^{-2}$}\\
\hline

$E_{M,L}$ & {\footnotesize $2.1\times 10^{-8} $} & {\footnotesize $6.0\times 10^{-6} $} &
{\footnotesize $2.7\times 10^{-2} $} & {\footnotesize $1.2\times 10^{-11}$}& {\footnotesize $6.5\times 10^{-6}$} & {\footnotesize $2.7\times 10^{-2}$}& {\footnotesize $2.4\times 10^{-14}$}& {\footnotesize $6.6\times 10^{-6}$}&{\footnotesize $2.7\times 10^{-2}$}\\
\hline

\multirow{3}{*}{$E_{M,S}$} & {\footnotesize $5.6\times 10^{-2} $} & {\footnotesize $7.7\times 10^{-5} $} &{\footnotesize $5.7\times 10^{-3} $} & {\footnotesize $5.5\times 10^{-2}$}& {\footnotesize $7.5\times 10^{-5}$} & {\footnotesize $5.8\times 10^{-3}$}& {\footnotesize $1.7\times 10^{-8}$}& {\footnotesize $4.3\times 10^{-5}$}&{\footnotesize $6.1\times 10^{-3}$}\\

& ({\footnotesize $5.6\times 10^{-2} $}) & ({\footnotesize $2.5\times 10^{-5} $}) & ({\footnotesize $1.2\times 10^{-3} $}) & ({\footnotesize $5.5\times 10^{-2}$})& ({\footnotesize $2.5\times 10^{-5}$}) & ({\footnotesize $1.1\times 10^{-3}$})& ({\footnotesize $1.7\times 10^{-8}$})& ({\footnotesize $1.5\times 10^{-5}$})&({\footnotesize $1.3\times 10^{-3}$})\\

& (100\%) & (32\%) & (21\%) & (100\%) &  (33\%) & (19\%) & (100\%) & ($\sim$35\%) & (21\%)\\
\hline

$E_{K,L}$ & {\footnotesize $2.8\times 10^{-6} $} & {\footnotesize $1.7\times 10^{-4} $} &
{\footnotesize $1.1\times 10^{-4} $} & {\footnotesize $1.6\times 10^{-8}$}& {\footnotesize $1.7\times 10^{-4}$} & {\footnotesize $8.6\times 10^{-5}$}& {\footnotesize $3.3\times 10^{-10}$}& {\footnotesize $1.9\times 10^{-4}$}&{\footnotesize $4.3\times 10^{-5}$}\\
\hline

\multirow{3}{*}{$E_{K,S}$} & {\footnotesize $6.5\times 10^{-3} $} & {\footnotesize $2.1\times 10^{-2} $} & {\footnotesize $9.7\times 10^{-3} $} & {\footnotesize $7.7\times 10^{-7}$}& {\footnotesize $2.3\times 10^{-2}$} & {\footnotesize $9.5\times 10^{-3}$}& {\footnotesize $1.3\times 10^{-3}$}& {\footnotesize $2.1\times 10^{-2}$}&{\footnotesize $1.1\times 10^{-2}$}\\

& ({\footnotesize $6.3\times 10^{-3} $}) & ({\footnotesize $1.8\times 10^{-4} $}) & ({\footnotesize $8.6\times 10^{-4} $}) & ({\footnotesize $6.8\times 10^{-7}$})& ({\footnotesize $2.7\times 10^{-4}$}) & ({\footnotesize $9.2\times 10^{-4}$})& ({\footnotesize $2.7\times 10^{-8}$})& ({\footnotesize $2.4\times 10^{-4}$})&({\footnotesize $8.5\times 10^{-4}$})\\

& (97\%) & (1\%) & (9\%) & (88\%) &  (1\%) & (10\%) & ($\sim0$\%) & (1\%) & (8\%)\\

\hline

\end{tabular}

\caption{Large scale: $k=1$, forcing scale: $k=2\sim 6$, smaller scale: $k=7\sim k_{max}$(quantity in parentheses). $E_{M,s}(t)=\sum_{k=2}^{k_{max}}E_{M,s}(k, t)$. The percentage is the ratio of smaller scale to small scale:$\sim\sum_{k=7}^{k_{max}}/\sum_{k=2}^{k_{max}}$.}
\end{table*}

\begin{figure*}
\centering{
     \subfigure[]{
     \includegraphics[width=8.55cm]{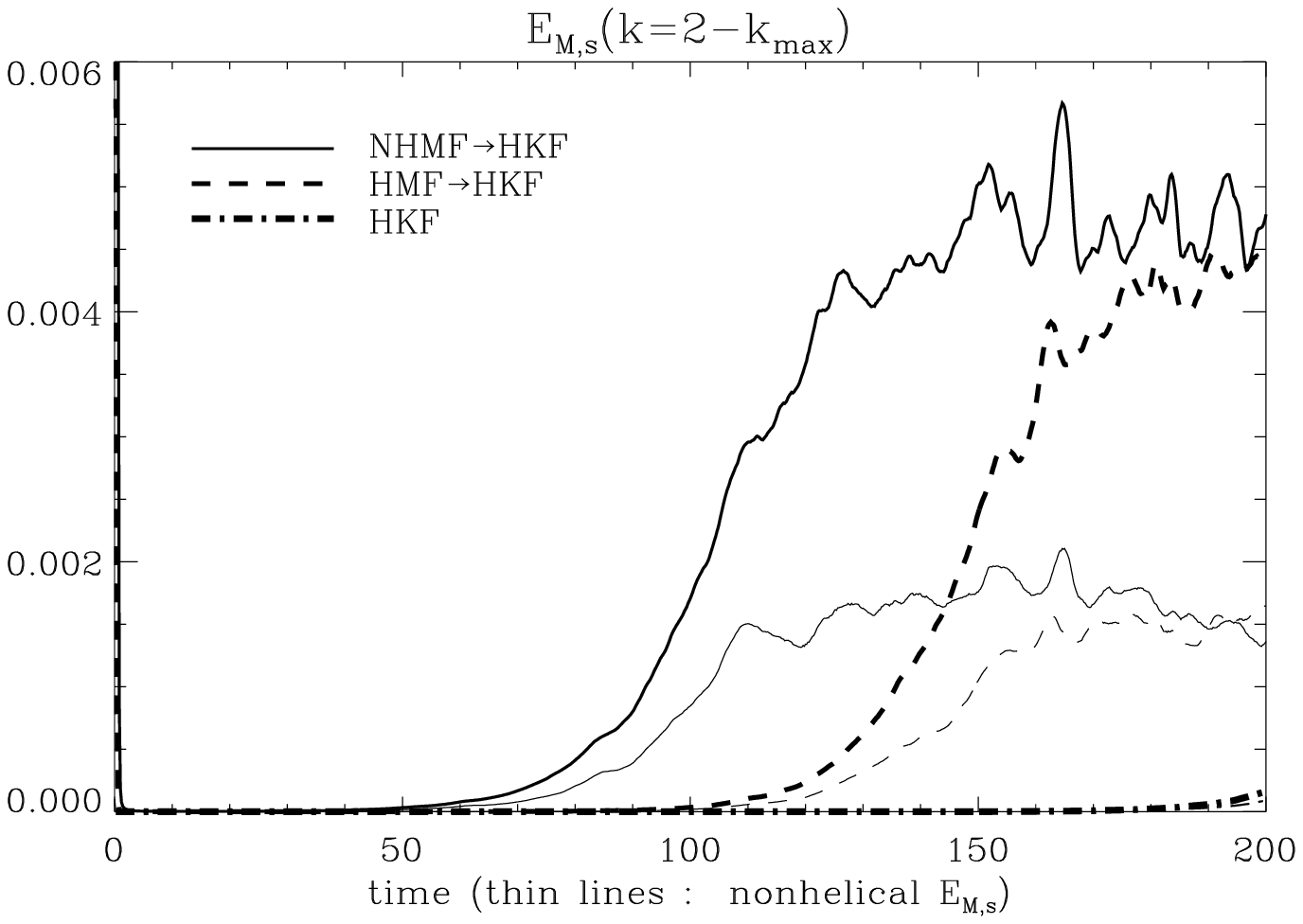}
     \label{1a}}
     \subfigure[]{
     \includegraphics[width=8.5cm]{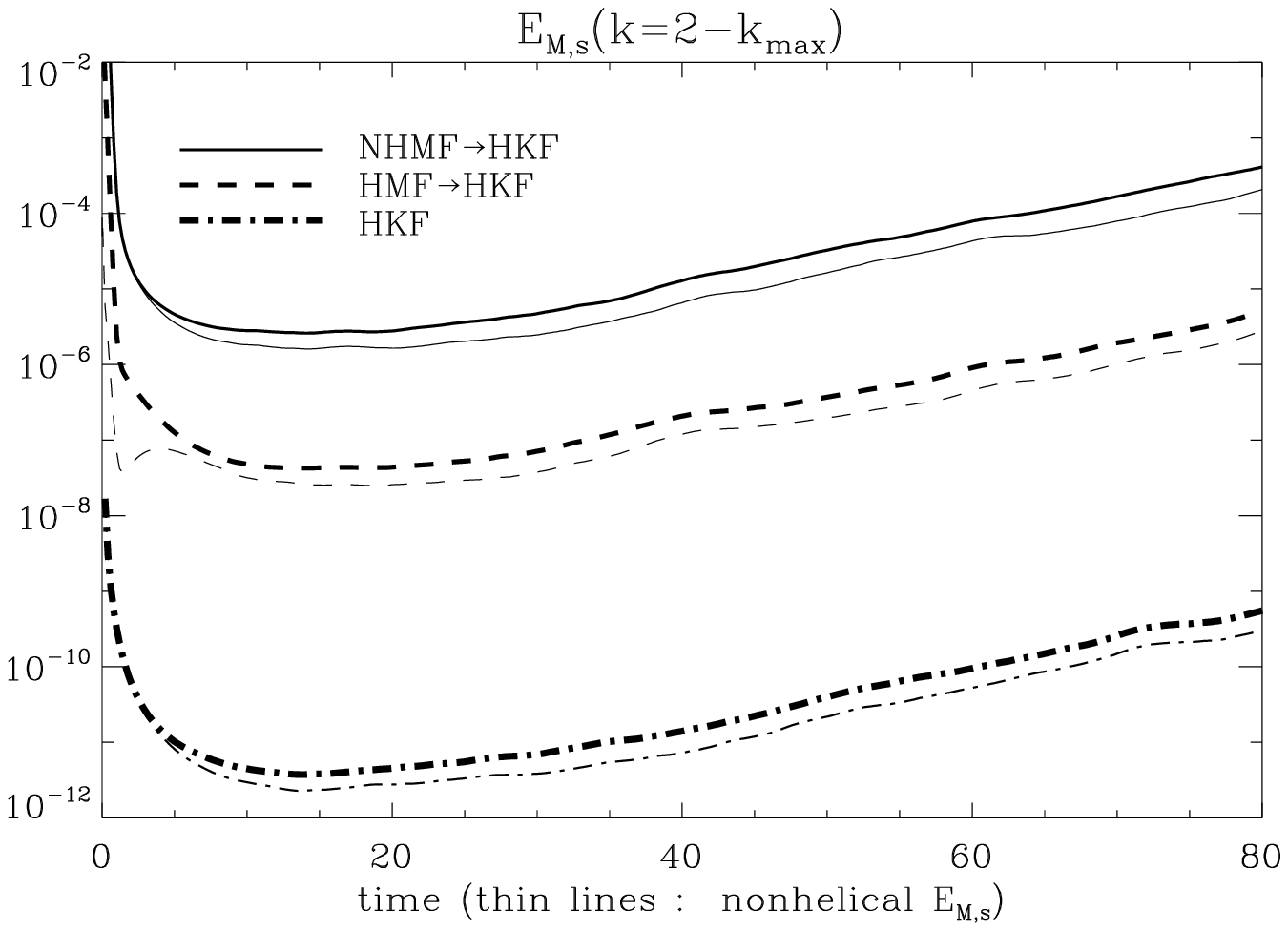}
     \label{1b}}
   \subfigure[]{
     \includegraphics[width=8.2cm]{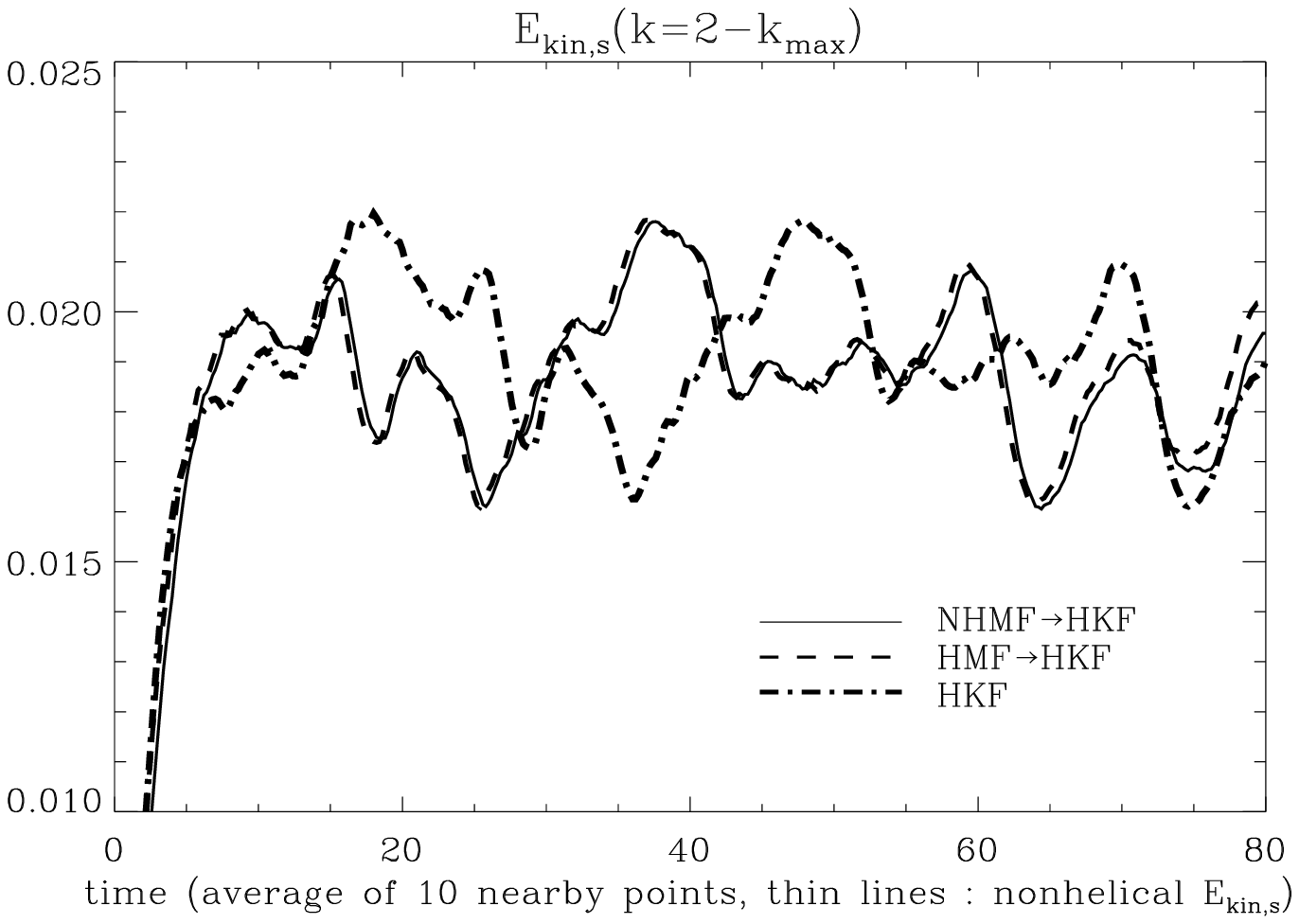}
     \label{1c}}
   \subfigure[]{
     \includegraphics[width=8.2cm]{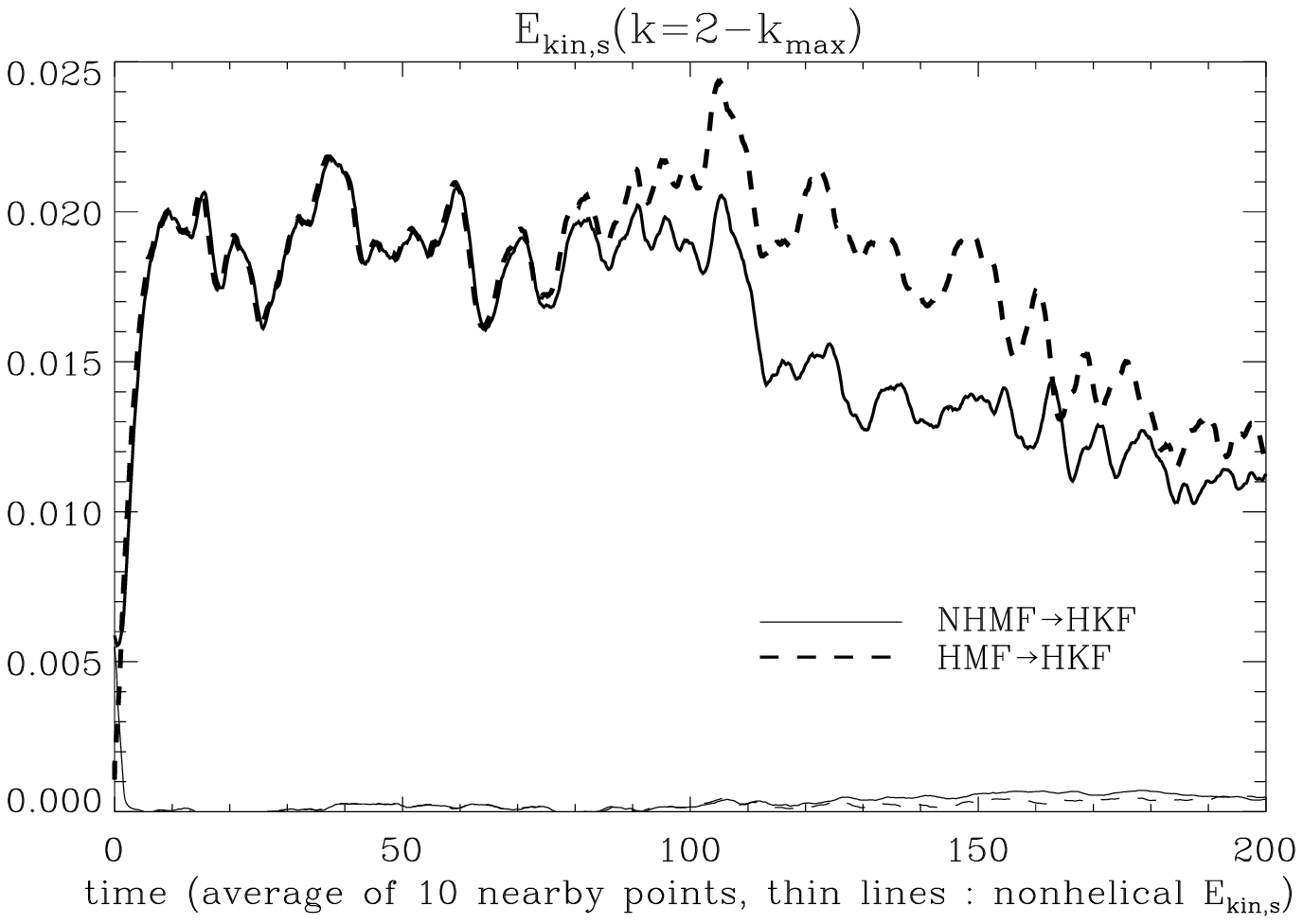}
     \label{1d}}
   \subfigure[]{
     \includegraphics[width=8.3cm]{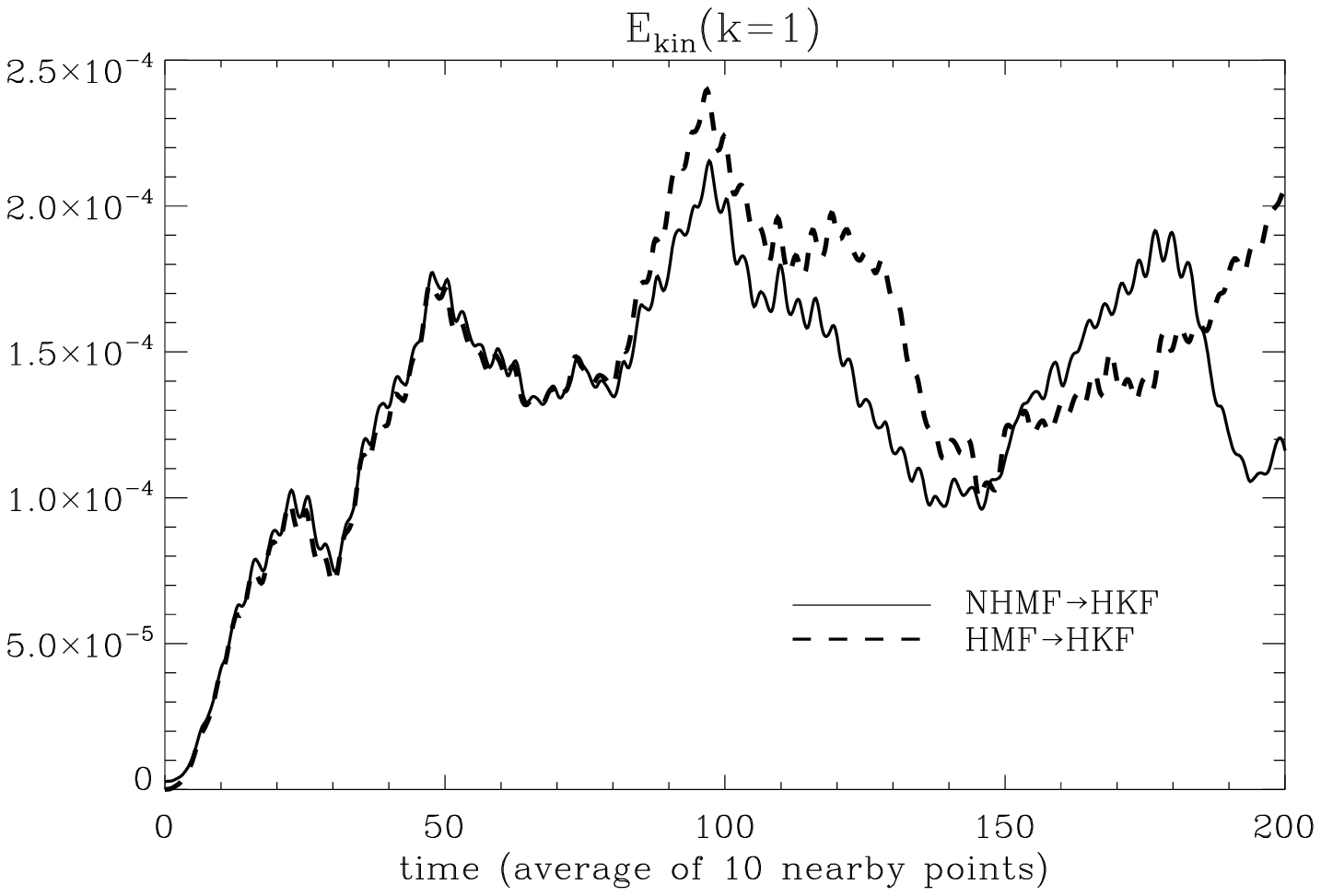}
     \label{1e}}
   \subfigure[]{
     \includegraphics[width=8.3cm]{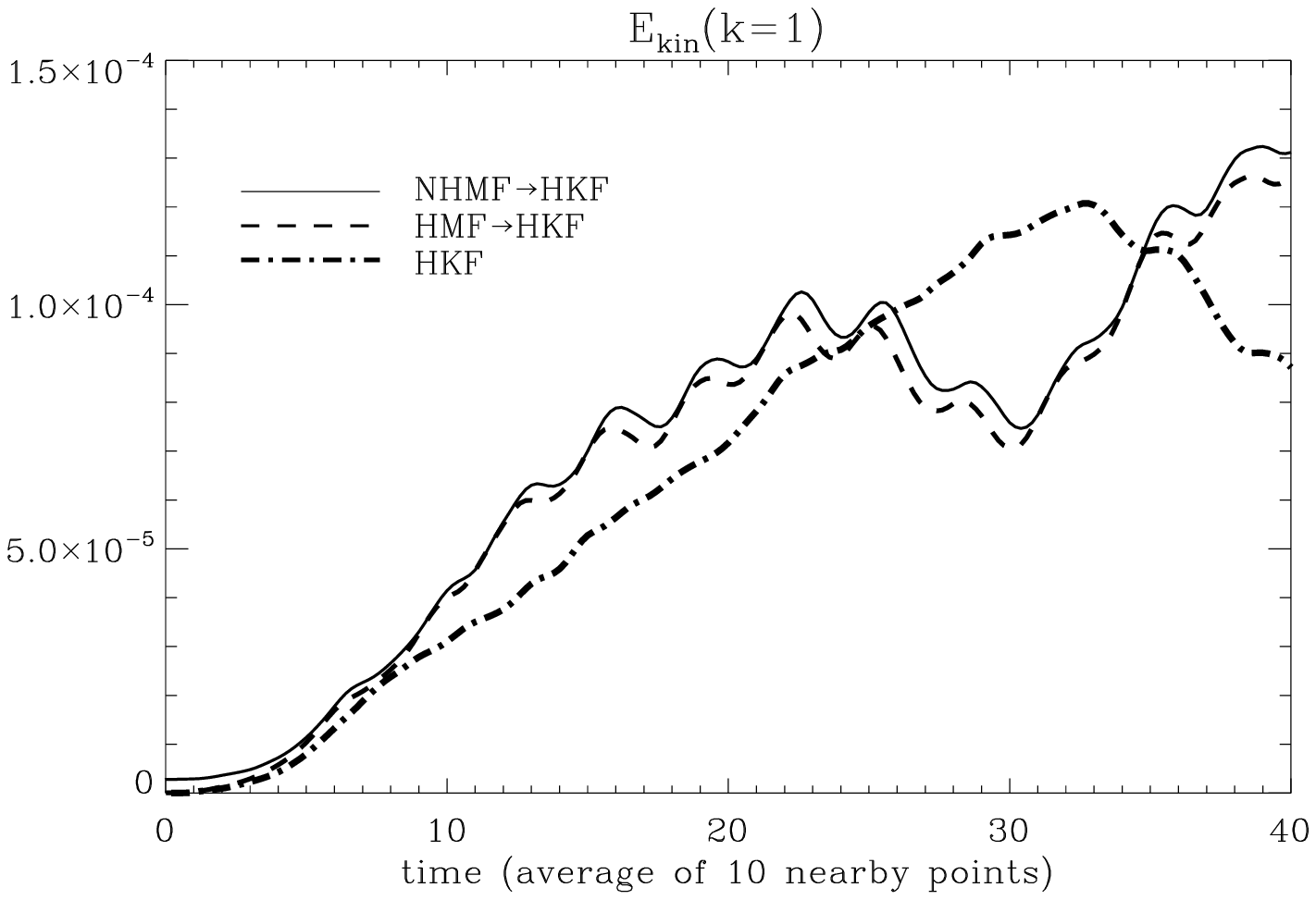}
     \label{1f}}
\caption{Preliminary simulation effect is removed by shifting the time unit(1st simulation: -10.6, 2nd simulation: -13.0). (a) $E_{M0,s}$: $5.6\times 10^{-2}$(1st), $5.5\times 10^{-2}$(2nd), $1.7\times 10^{-8}$(3rd). (c), (f) In the very early time regime, all $E_{kin}$s evolve in the same way being independent of the initial values. (d) $E_{kin}$ of smaller scale eddy branches off earlier.}
}
\end{figure*}

\begin{figure*}
\centering{
   \subfigure[]{
     \includegraphics[width=8.2cm]{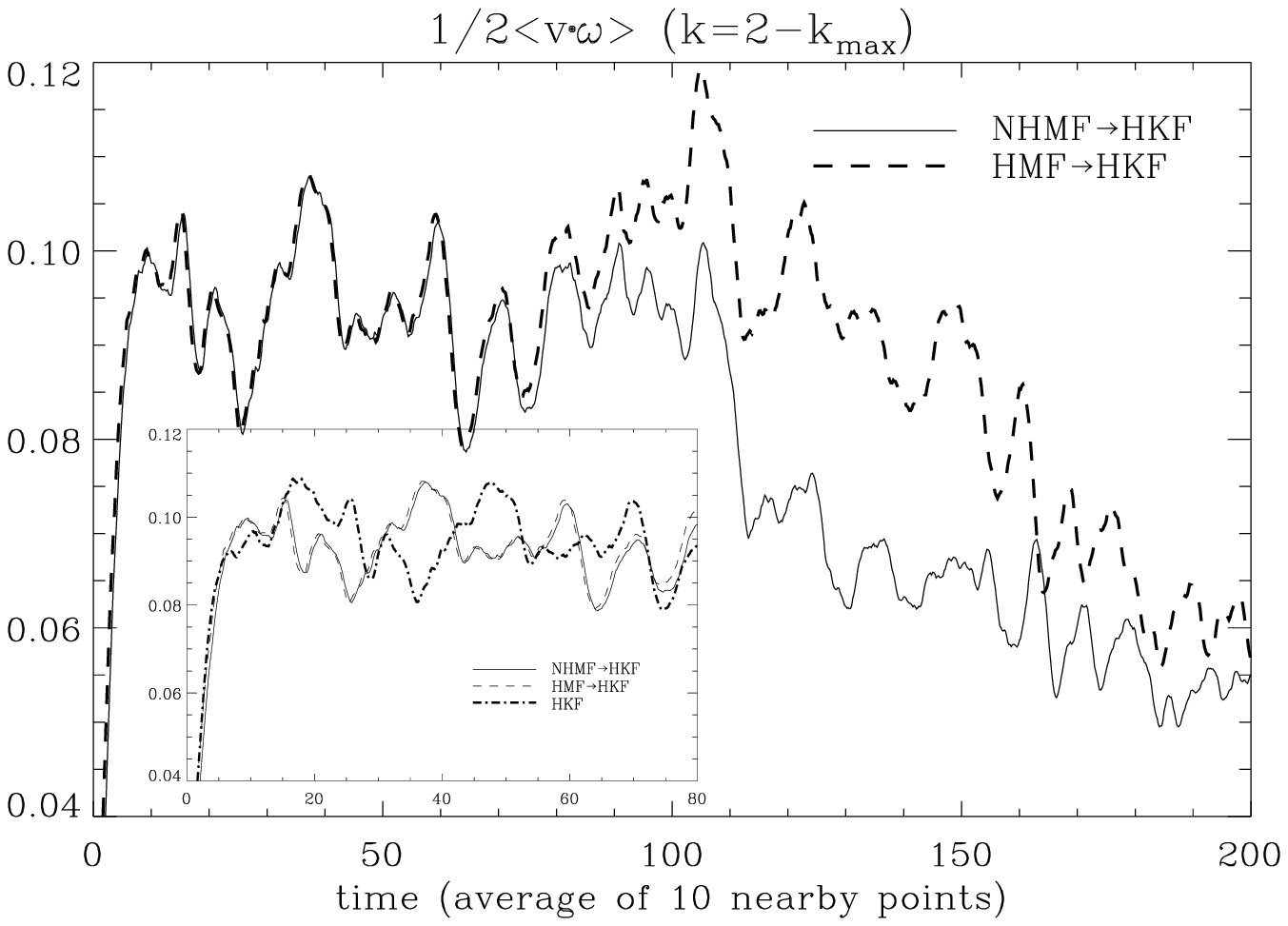}
     \label{2a}}
   \subfigure[]{
     \includegraphics[width=8.3cm]{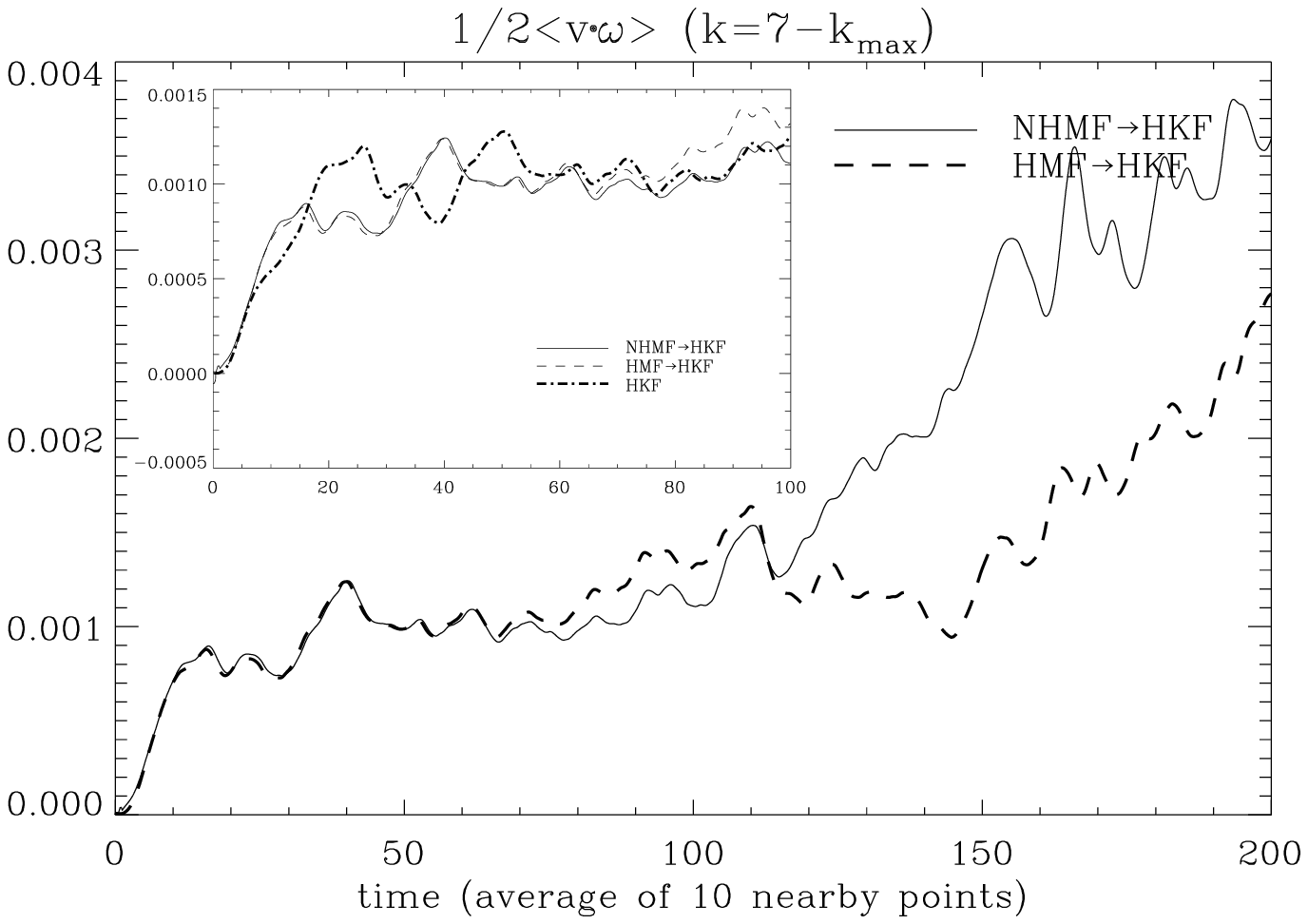}
     \label{2b}}
     \subfigure[]{
     \includegraphics[width=8.3cm]{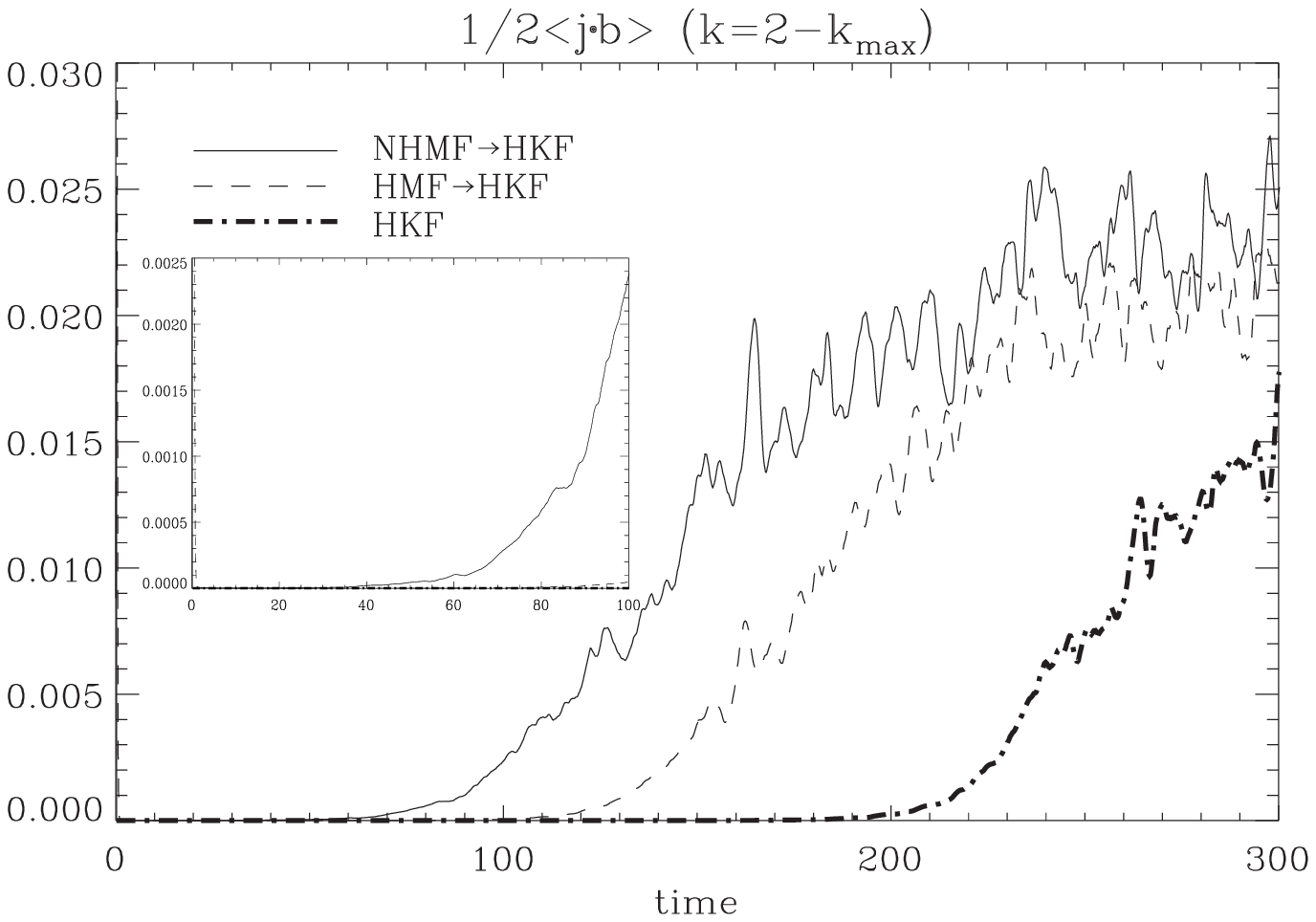}
     \label{2c}}
   \subfigure[]{
     \includegraphics[width=8.3cm]{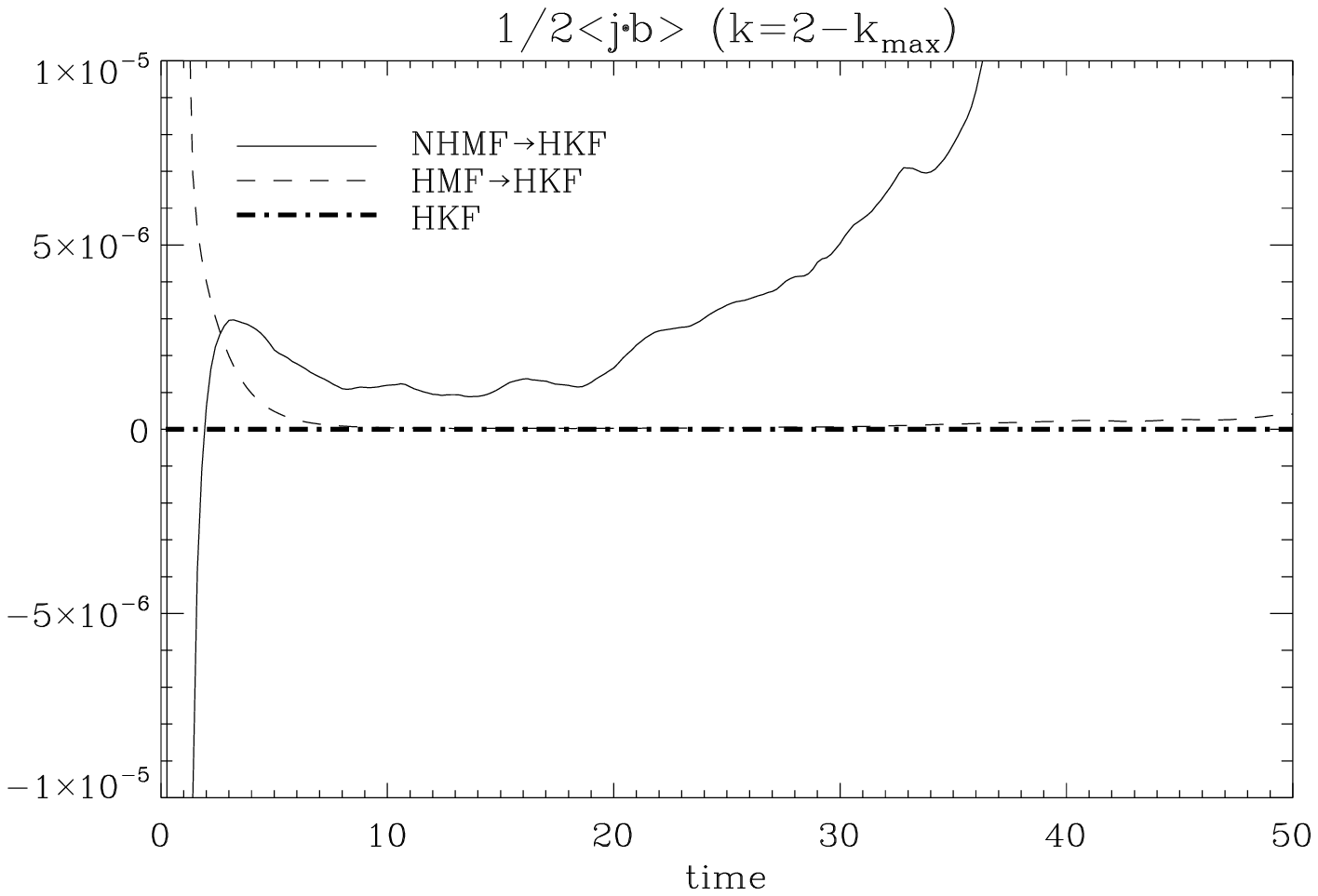}
     \label{2d}}
     \subfigure[]{
     \includegraphics[width=8.3cm]{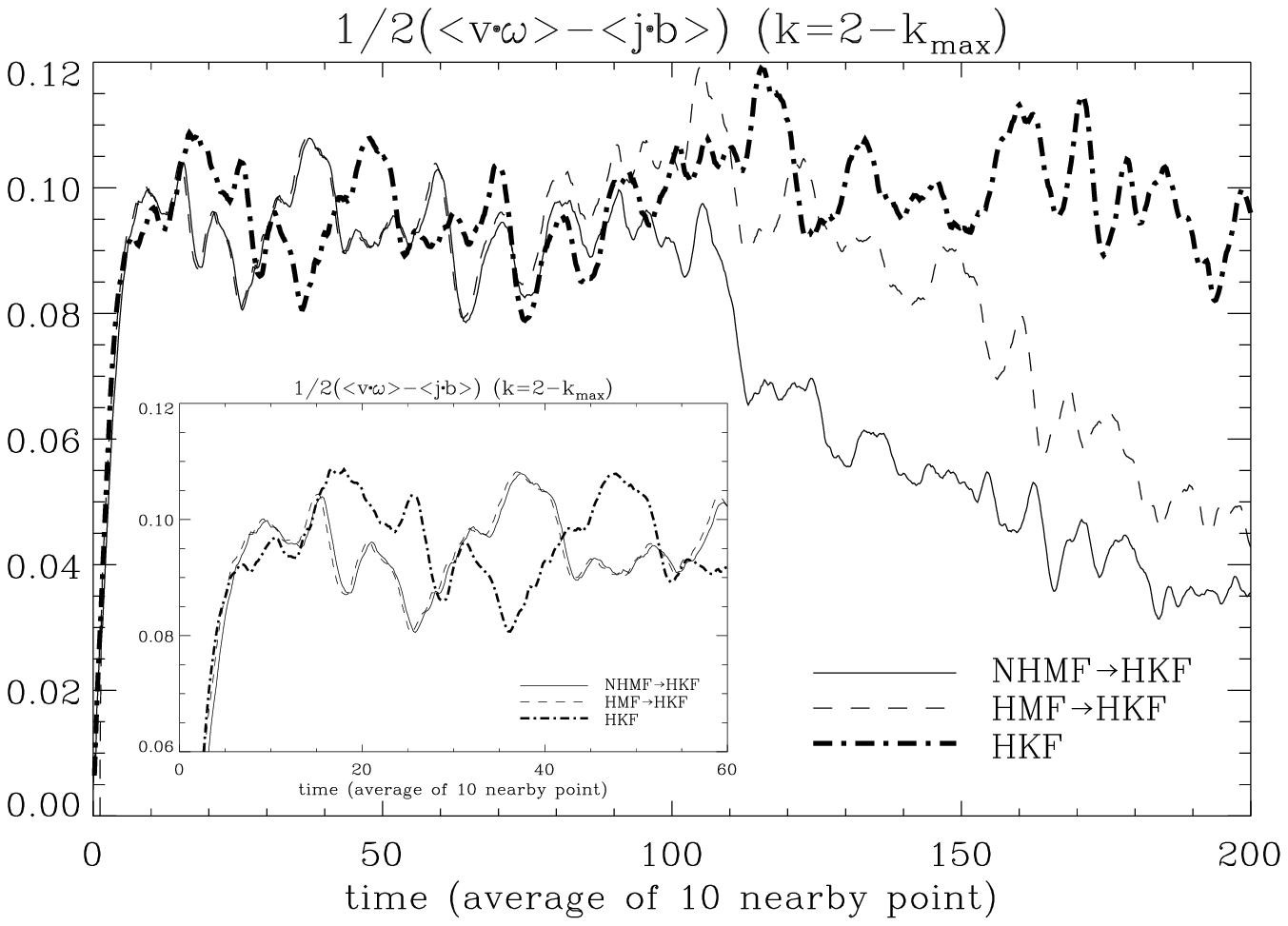}
     \label{2e}}
   \subfigure[]{
     \includegraphics[width=8.5cm]{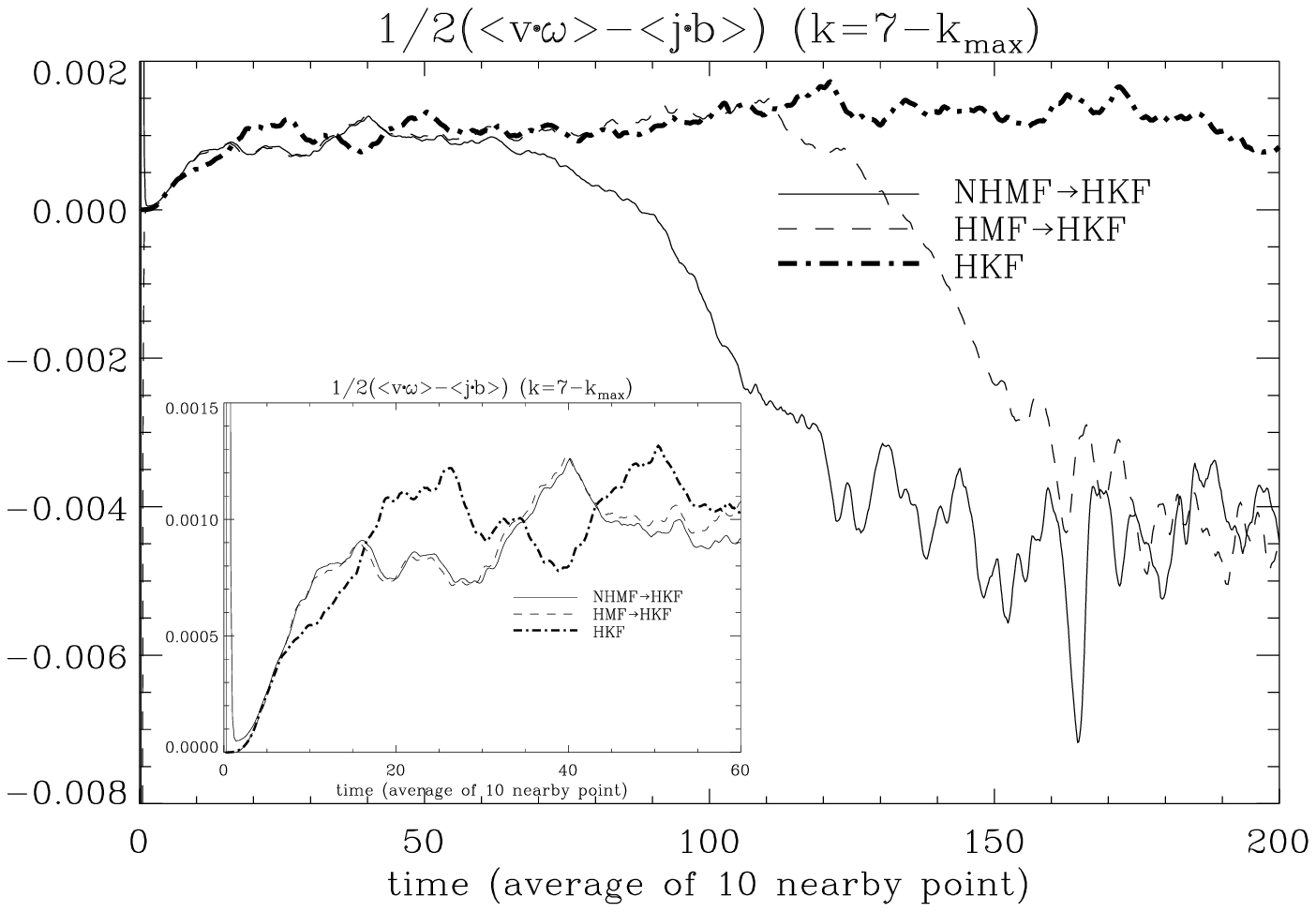}
     \label{2f}}
\caption{Small box includes $\langle v\cdot \omega \rangle$ and residual helicity for 3rd simulation. Their profiles are influenced by the large scale field of the 3rd simulation and take different routes unlike 1st and 2nd simulation. (d) The evolution of $\langle {\bf j}\cdot {\bf b}\rangle$ needs to be compared with that of $E_{M,s}$ in Fig.\ref{1b}.}}
\end{figure*}

\begin{figure*}
\centering{
   \subfigure[]{
     \includegraphics[width=8.4cm]{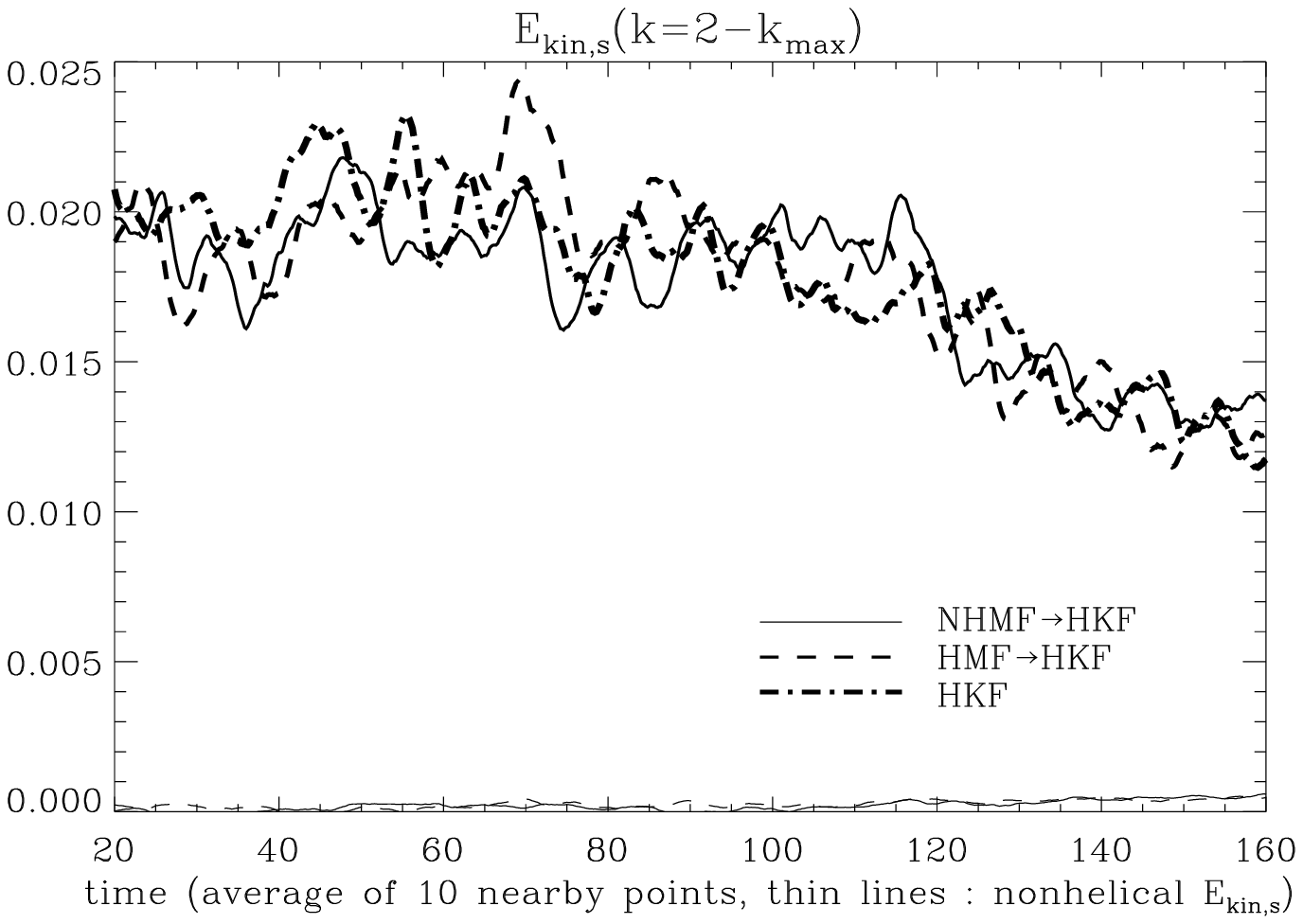}
     \label{3a}}
   \subfigure[]{
     \includegraphics[width=8.4cm]{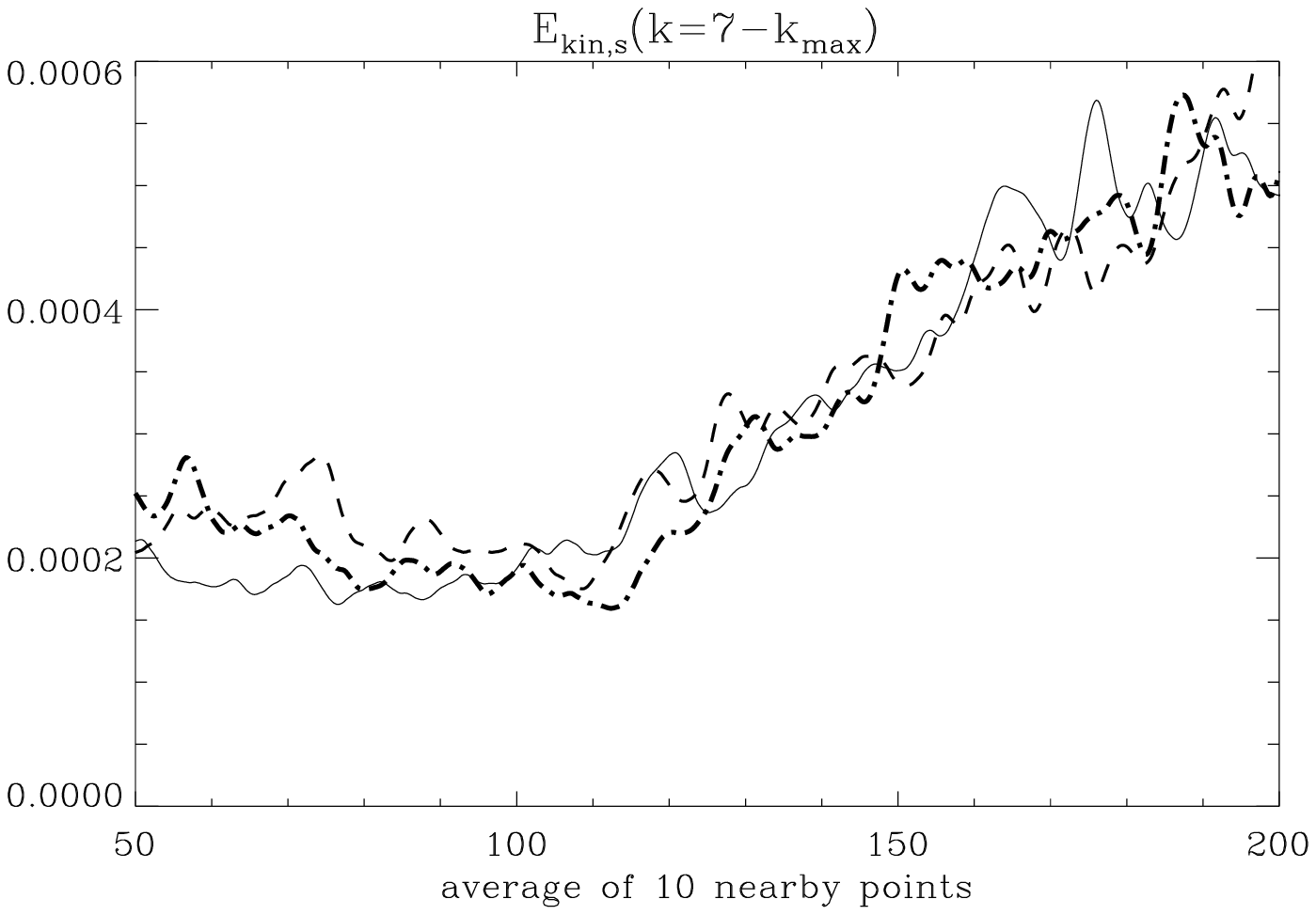}
     \label{3b}}
   \subfigure[]{
     \includegraphics[width=8.4cm]{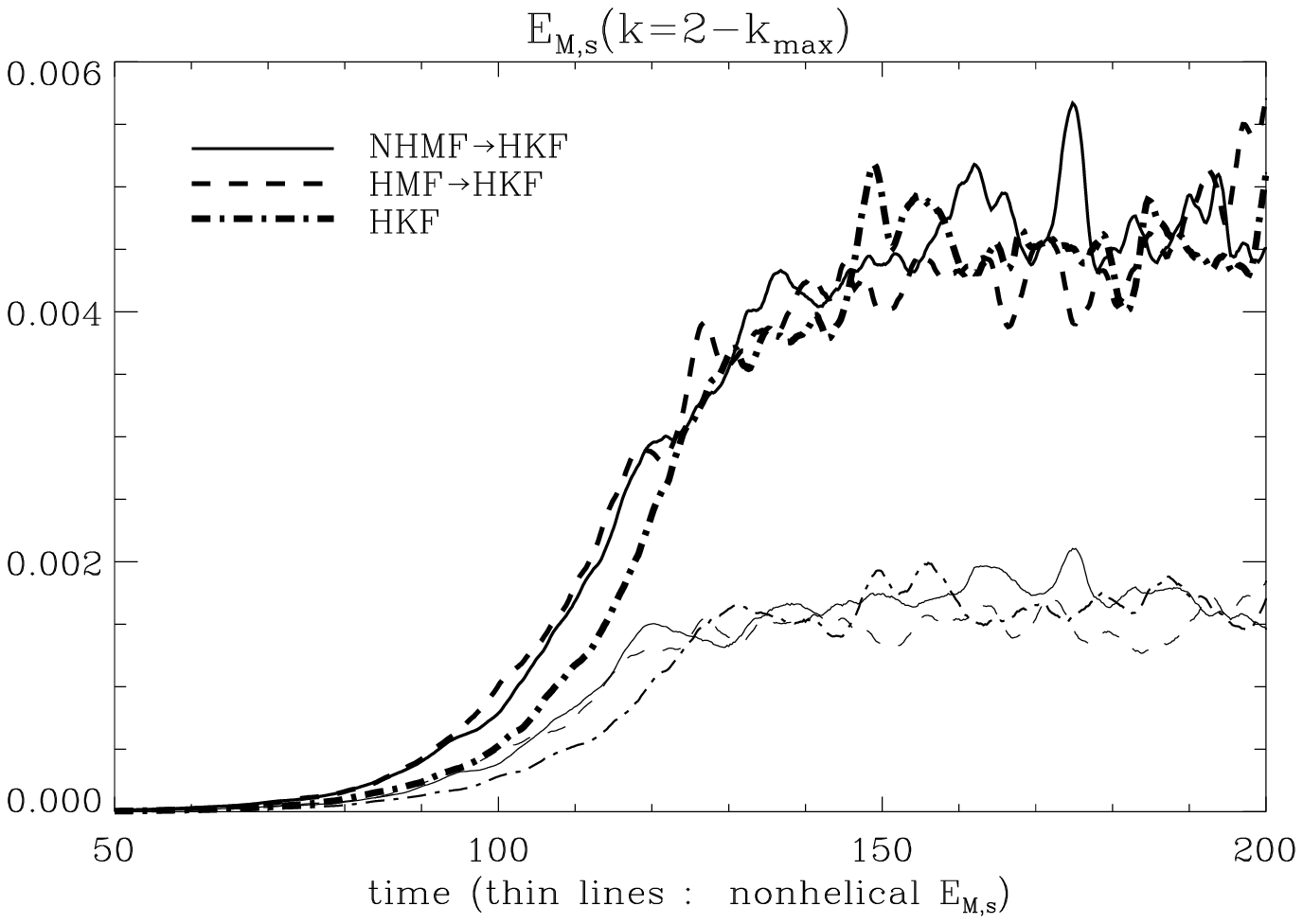}
     \label{3c}}
   \subfigure[]{
     \includegraphics[width=8.4cm]{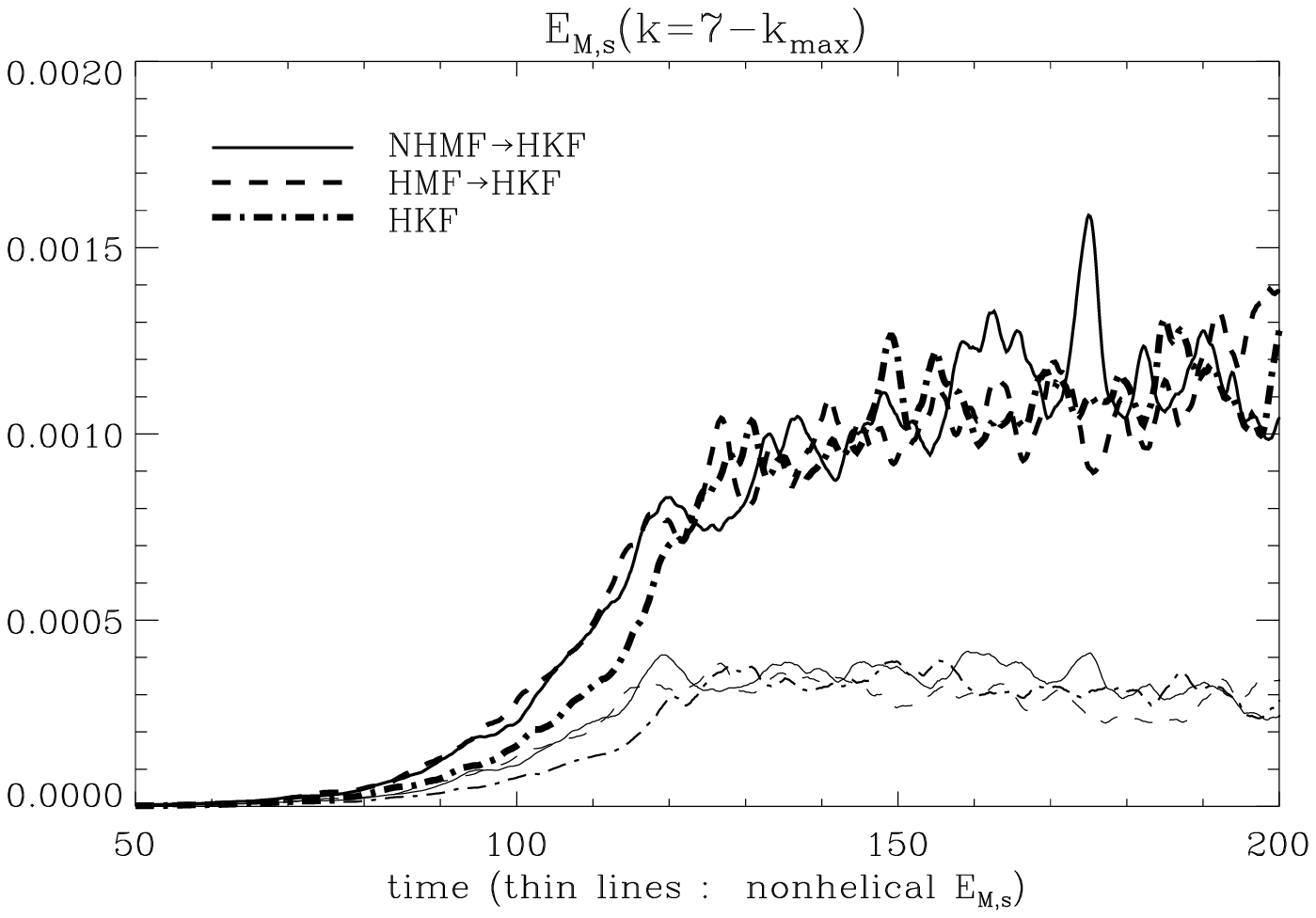}
     \label{3d}}
     \subfigure[]{
     \includegraphics[width=8.4cm]{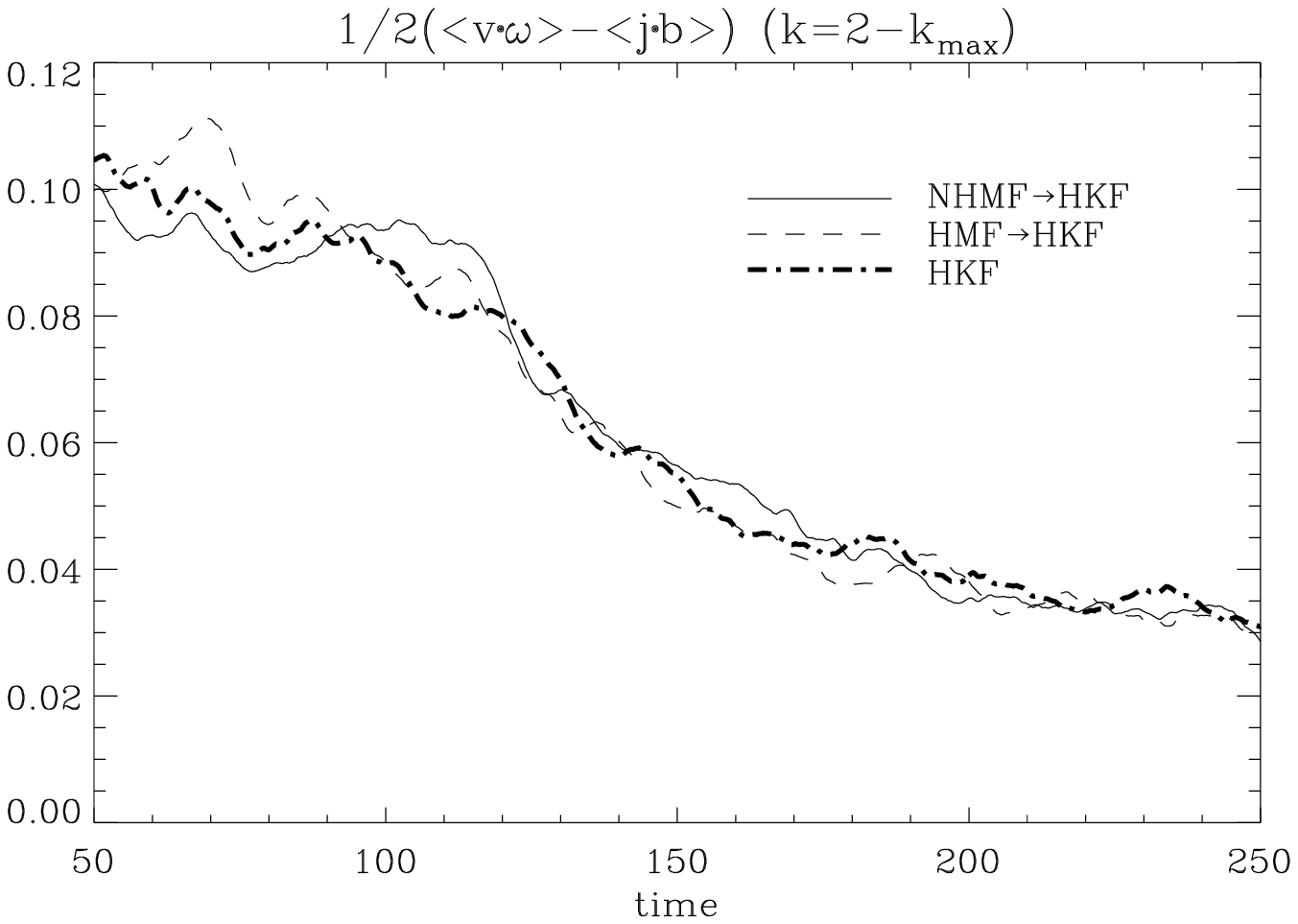}
     \label{3e}}
   \subfigure[]{
     \includegraphics[width=8.4cm]{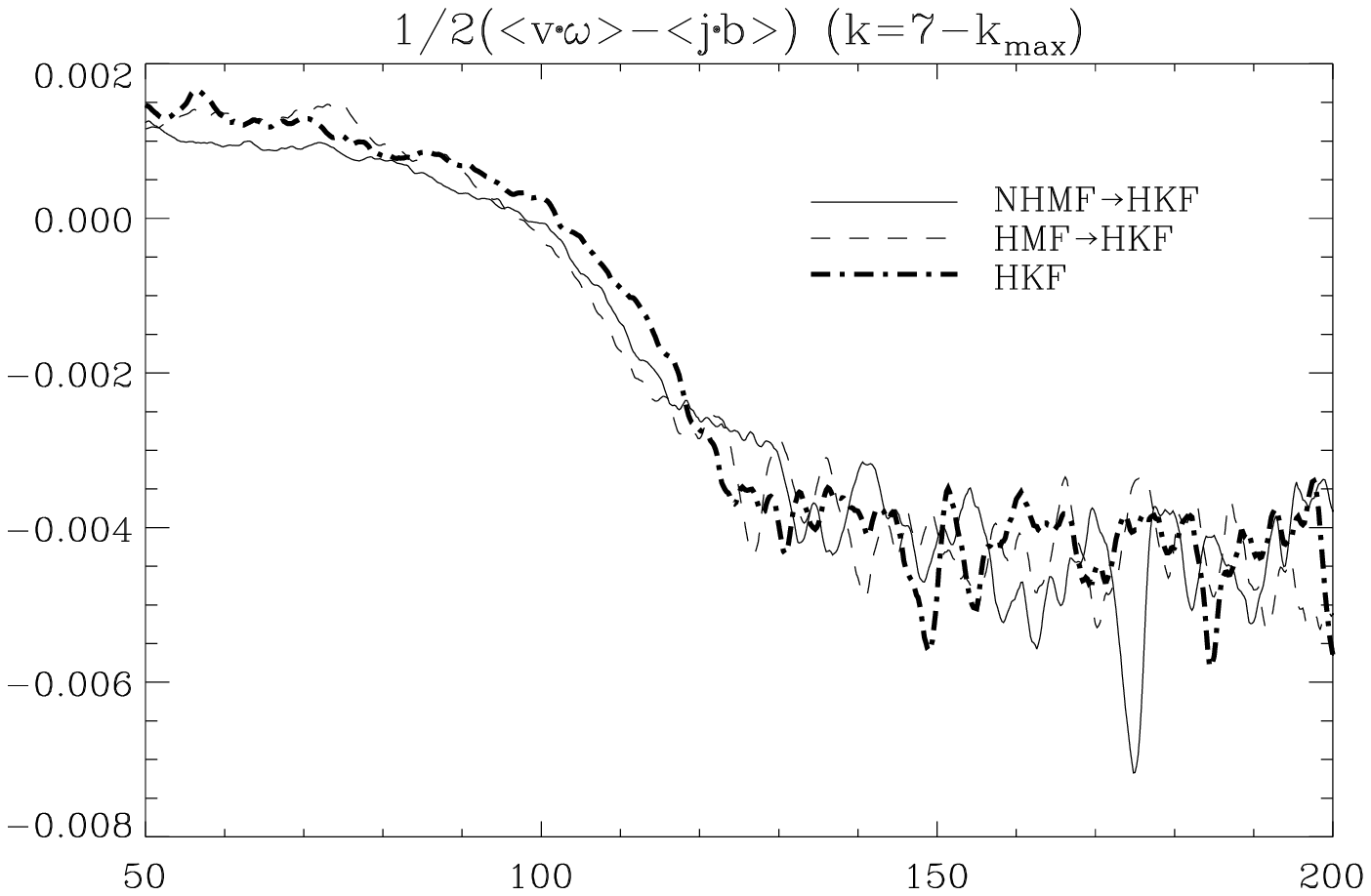}
     \label{3f}}
\caption{The plots of 2nd and 3rd are shifted by -50 and -116 for the comparison with 1st simulation after the onset.}
}
\end{figure*}

\section{Simulation result 2}
Table.1 provides information on the energy distributions of the evolving variables at each stage.  When $E_M$ or $H_M$ is about to rise, most $E_{kin}$ is located in the forcing scale regime($k=2\sim 6$) regardless of its initial distribution. The ratio of smaller scale $E_{kin}$($k=7\sim k_{max}$) to $E_{kin}$ of the whole small scale regime($k=2\sim k_{max}$) is about 1\%. After the onset, as large scale $E_M$ or $H_M$ grows, kinetic energy migrates towards the smaller scale. At this time the saturated ratio elevates up to $8\sim 10$\%. When large scale eddy needs more $H_M$ or $E_M$(onset position), more (helical) kinetic energy is located in the forcing scale. And if the inverse cascade of $H_M$ is less required, $E_{kin}$ in the forcing scale decreases and moves toward the smaller scale which has less helical effect but more dissipative effect. $E_{kin, s}$, more exactly $\langle {\bf v}\cdot {\bf\omega} \rangle$ plays the role of balancing the growth of large scale magnetic field. However, the evolution of $E_{kin,L}$ shows rather an irregular feature. In the early time regime $E_{kin,L}$ leads the growth of $E_{M,L}$ which generates $H_M$(Fig.1, 2, and 3). According to EDQNM approximation, the role of $E_{kin,L}$ with $E_{M,L}$ is related to the self distortion effect(the eddy damping rate $\mu_k$, Eq.(\ref{kinetic nu theta mu definition})). Smaller magnitude of $E_{kin,L}$ decreases $\mu_k$, which increases $\alpha$ effect and dissipation at the same time. However, more detailed simulation is necessary to check these theoretical inference. \\

\noindent Analytic equation like EQDNM or mean field dynamo theory does not explicitly explain the role of magnetic energy in the small scale. But simulation results provide some clues to the influence of $E_{M,s}$ on the dynamo. The ratio of smaller scale $E_M$ to that of the whole small scale is consistently regular, i.e., from onset: $\sim 30\%$ to saturation: $\sim 20\%$. The distribution of $E_{M,s}$, more exactly $\langle {\bf j}\cdot {\bf b}\rangle$, is related with the inverse cascade of $H_M$(or $E_M$) and balancing the growth rate of large scale magnetic field. $\langle j\cdot b\rangle$ in $\alpha$ coefficient does not always quench the large scale magnetic field. As Fig.\ref{18}, \ref{19} show, when the necessity of inverse cascade of magnetic energy is large, forcing scale $H_M$ is negative so that $\alpha$ effect is enhanced(the kinetic helicity in $\alpha$ coefficient keeps positive). As the large scale field saturates, the sign of forcing scale magnetic helicity grows to be positive, i.e., lowering $\alpha$ effect.\\

\noindent With more detailed plots we can investigate the dynamic properties of small scale regime with respect to the large scale B field growth.\\
Fig.\ref{1a} shows the evolution of small scale $E_{M,s}$(=$\sum_{k=2}^{k=k_{max}}E_M(k)$) and nonhelical $E_{M,s}$($=E_{M,s}-kH_{M,s}$, thin line). To remove the preliminary simulation, $NHMF\rightarrow HKF$($\equiv$ 1st simulation from now on) was shifted by -$10.6$ time unit using $t'\equiv t-10.6$ and $HMF\rightarrow HKF$($\equiv$ 2nd simulation) was shifted by -$13.0$ time unit using $t'\equiv t-13.0$. But $HKF$($\equiv$ 3rd simulation) was not shifted. The difference between the thick and thin line is the helical B-field, which is generated counteractively as the large scale helical magnetic field grows. This helical magnetic field in the small scale can induce the growth of large scale B-field($\langle j\cdot b \rangle>0$) or suppress it($\langle j\cdot b \rangle<0$).\\

\noindent Fig.\ref{1b} shows the profiles of small scale $E_M$ in the early time regime. $E_M$ of the 1st simulation is relatively larger than that of other cases, and $E_M$ of the 3rd simulation is the least. The initial small scale $E_{M0,s}$($5.6\times10^{-2}$ of the 1st simulation, $5.5\times10^{-2}$ of the 2nd simulation) due to the preliminary simulation drops till $t\sim 10$ and begins to grow again. The different minimum value and evolution of each field profile imply some important clues to the relation between large scale magnetic field and small scale magnetic field. In addition, the origin of helical magnetic field can be inferred from Fig.\ref{1a}, \ref{1b}.\\

\noindent The magnetic helicity is the topological linking number of magnetic fields. But statistically it can be considered as the correlation between different components of magnetic field(\cite{2011Springer.book.....Y}).
\begin{eqnarray}
\langle B_i(k)B_j(-k)\rangle =P_{ij}(k)\frac{E_M(k)}{4\pi k^2}+\frac{i}{2}\frac{k_l}{k^2}\epsilon_{ijl}H_M(k)\\
\bigg(P_{ij}(k)=\delta_{ij}-\frac{k_ik_j}{k^2}\bigg).\nonumber
\label{BiBj correlation}
\end{eqnarray}
Since magnetic helicity $\langle a\cdot b \rangle$ cannot be larger than $2E_M/k$. Small scale magnetic energy has a lower bound proportional to the small scale magnetic helicity. In addition, the growth of small scale magnetic helicity depends on that of large scale magnetic helicity in terms of the conservation of magnetic helicity in the system. All of these explain the reasons of quick drop of $E_{M0}$ and different evolution of $E_M$ in the small scale. The role of nonhelical $E_M$ becomes clear with the comparison of $E_{kin}$. We will discuss about this again. The initial B-field plays the role of seed field in MHD dynamo, and at the same time the correlation between its different components constrains the growth of large scale magnetic field dynamically changing the sign and magnitude.\\

\noindent Fig.\ref{1c}, \ref{1d} show $E_{kin}$ spectra of the 1st and 2nd simulation are very similar. Fig.\ref{1b}, \ref{1c}, and \ref{1d} imply the profile of kinetic energy does not depend on the small scale $E_M$ much as long as $E_M$ is not too much different. Fig.\ref{1e} and \ref{1f} also show large scale $E_{kin}$ is independent of $E_M$ when the magnetic energy is not significantly different. In addition, Fig.\ref{1d}, \ref{1e} clearly show that $E_{kin}$ drops when large scale $E_M$ begins to rise, i.e., onset position. At the onset point of the 1st simulation, $t\sim70$, $E_{kin}$ of this simulation begins to drop. And around $t\sim 120$, onset position of the 2nd simulation, $E_{kin}$ of this simulation also begins to drop. They meet again each other when large $E_M$ of each simulation gets saturated.\\

\noindent In Navier Stokes equation(Eq.2), Lorentz force(${\bf J}\times {\bf B}$) can be decomposed into magnetic tension(${\bf B}$$\cdot$$\nabla$${\bf B}$) and pressure($-\nabla B^2/2$). The force parallel to B-field from magnetic tension and pressure is canceled out. Only the force perpendicular to the magnetic field line like $(B^2/R_c-\nabla(B^2/2))\hat{n}$, $R_c$: radius of curvature, \cite{2003dysu.book..217P}) exists, as the definition of Lorentz force implies. When the growth of $E_M$ accelerates near the onset point, the compressive force($-\nabla B^2/2$) normal to B-field grows so that the net effect of Lorentz force becomes negative. This presses the plasma and causes the geometrical changes of magnetic fields. The kinetic motion of plasma slows down.\\

\noindent Figure.\ref{1c}, \ref{1f} also show there is a time regime($t<7\sim10$) where the profile of evolving $E_{kin}$ is independent of the initial values. This occurs when Lorentz force is still weak, and looks like the corresponding concept of the kinematic regime.\\

\noindent Fig.\ref{2a}$\sim$\ref{2f} include kinetic and current helicity. Kinetic helicity profiles shown in Fig.\ref{2a} are very similar to those of kinetic energy. In addition, Fig.\ref{2b} shows the clear migration of kinetic helicity toward the smaller scale regime. Fig.\ref{2c}, \ref{2d} are the current helicity in small scale regime. Especially the profile of evolving current helicity in Fig.\ref{2d} suggests that the evolution of $E_M$ in small scale be determined by the growth of large scale magnetic energy(helicity). As mentioned, small scale $E_{M0}$ of the 1st and 2nd simulation are almost the same. We know the different large scale $E_{M0}$(or $H_{M0}$) causes the different growth rates($\partial \overline{{\bf B}}/\partial t \sim \langle {\bf v}\cdot {\bf \omega} \rangle \overline{{\bf B}}$). The fast growth of negative $H_{M}$ in the large scale requires the fast growth of positive $H_{M}$ in small scale to conserve $H_{M}$ in the system. The evolution of small scale $E_M$ or $H_M$ is highly influenced by the initial large scale $E_M$ or $H_M$. The quick change of sign with the sequent fast growth of small scale $\langle {\bf j}\cdot {\bf b} \rangle$ of the 1st simulation and the fast decay of small scale $E_M$ of the second simulation support this fact very well.\\

\noindent Fig.\ref{2e}, \ref{2f} show how residual helicity evolves. In the very early time regime the residual helicity does not depend on the initial conditions. This phenomenon is an inevitable consequence for the kinetically driven MHD dynamo.\\

\noindent In Fig.\ref{3a}$\sim$\ref{3f}, the onset positions of 2nd and 3rd simulation were shifted by $t\rightarrow t-(120-70)$ and $t\rightarrow t-(186-70)$ to compare the behaviors of field profiles after the onset. Except some minor differences due to the turbulene, all three simulations have the same field profiles. This indicates the saturation of MHD dynamo is independent of the initial conditions.\\

\noindent On the other hand in Fig.\ref{10}$\sim$\ref{12}, we pointed out the cascade of kinetic energy toward larger scale. This looks to contradict the established theory that the inverse cascade occurs with the strong rotation effect or in the ideal (quasi) two dimensional hydrodynamic system. This conclusion is based on the conservation of physical quantities like energy and enstrophy. The modified expression of $\langle k \rangle$(\cite{2004tise.book.....D}) is,\\\\
\begin{eqnarray*}
\frac{d\langle k\rangle}{dt}&=&\frac{d}{dt}\bigg(\frac{\int_0^\infty k\,E_v(k,t)\,dk}{\int_0^\infty E_v(k,t) \,dk}\bigg)\nonumber \\
&=&\frac{-\frac{1}{2k_c}\frac{d}{dt}\big[\int_0^\infty (k-k_c)^2\,E_v(k,t) \,dk
]}{\int_0^\infty E_v(k,t) \,dk}\nonumber \\
&&+\frac{\frac{1}{2k_c}\frac{d}{dt}\big[\int_0^\infty (k^2+k_f^2)\,E_v(k,t) \,dk
\big]}{\int_0^\infty E_v(k,t) \,dk}\nonumber \\
&&-\frac{(\int_0^\infty k\,E_v(k,t)\,dk)\frac{d}{dt}\int_0^\infty E_v(k,t) \,dk}
{(\int_0^\infty E_v(k,t) \,dk)^2}.
\label{Inverse cascade}
\end{eqnarray*}
($k_c=\int kE(k,t)\,dk/\int E(k,t)\,dk$)
\\

\noindent If total energy $\int_0^\infty E_v(k,t)\,dk$ and enstropy $\int_0^\infty k^2\,E_v(k,t)\,dk$ are conserved, the first term in the right hand side determines $\langle k\rangle$. Since the usual spreading $E_v(k,t)$ in turbulent flow makes $\int_0^\infty (k-k_c)^2\,E_v(k,t) \,dk$ grow, $\langle k \rangle$ decreases(inverse cascade). However, if enstrophy or energy is not conserved(by the external source), $\langle k \rangle$ can grow or decrease according to $\dot{E}_v$ and $\langle \dot{\omega^2}/2 \rangle$. \cite{2012PhRvL.108p4501B} showed that the reverse cascade of hydrodynamic energy occurs when the mirror symmetry is broken(helicity). In addition, there were another trials to explain the forward cascade of $E_{kin}$ using the canonical ensemble average(\cite{2008matu.book.....B}, \cite{1975JFM....68..769F}). If $E_{tot}$, $H_M$, and $H_c$(=$\langle {\bf u}\cdot {\bf b}\rangle$) are conserved quantities(ideal three dimensional MHD system), the form of $E_{kin}$($\sim\langle u_iu_i \rangle$) calculated using the canonical partition function $Z^{-1}exp(-\alpha E_{tot}-\beta H_M-\gamma H_C)$ ($Z$: a normalization factor) shows forward cascade. However, if $E_{tot}(=E_M+E_{kin}$) is not conserved because of the external forcing or some other sources, another term should be added to $E_{tot}$ in the partition function; this changes the averaged $E_{kin}$ into a new form that allows the backward cascade.

\section{Conclusion}
\noindent Based on simulation and theory, we have investigated the influence of $IC$s, the role of (non)helical field with the pressure in the energy transfer.\\

\noindent The growth rate of large scale magnetic field is chiefly proportional to the large scale initial values. In contrast, its saturation depends on the external driving source and the intrinsic properties like $\eta$ or $\nu$ instead of $IC$s.\\

\noindent Comparing the simulation results, we have seen how the helical and nonhelical magnetic field constrain MHD dynamo. The helical magnetic field in the small scale has been thought to quench the growth of large scale magnetic field. As $\langle {\bf A}\cdot {\bf B}\rangle$ increases the opposite sign of small scale $\langle {\bf a}\cdot {\bf b}\rangle$ also grows. So the amplification of large scale magnetic field slows down and saturates. However, in the early time regime $\langle {\bf a}\cdot {\bf b}\rangle$ and large scale $H_M$ have the same sign; thus, $\langle {\bf a}\cdot {\bf b}\rangle$ boosts the growth of large scale magnetic field.\\

\noindent On the other hand, growing nonhelical magnetic field presses the plasma through Lorentz force(magnetic pressure) and slows down the motion, which constrains the magnetic fields eventually. Of course, the evolution of small scale fields is also influenced by the large scale $IC$s and the evolving large scale field.\\

\noindent Besides, it is observed kinetic energy migrates backward when the external energy flows into the three dimensional MHD system. And the velocity field in the early time regime seems to play a preceding role in the MHD dynamo.


\section{Acknowledgement}
Kiwan Park acknowledges support from US NSF grants PHY0903797, AST1109285, and a Horton Fellowship from the Laboratory for Laser Energetics at the University of Rochester.

\appendix
\section{Eddy Damped Quasi Normal Markovianization}
\subsection{Two point closure}
Navier Stokes equation for the incompressible fluid is
\begin{eqnarray}
\bigg(\frac{\partial}{\partial t}+\nu k^2\bigg)u_i(k)=-ik_mP_{ij}(k)\sum_{p+q=k}u_j(p)u_m(q)\sim \langle{\bf u}{\bf u}\rangle.
\label{Navier Stokes Fouier 1}
\end{eqnarray}
This equation requires information on the second order correlation equation. Then we need to solve another differential equation:
\begin{eqnarray}
\bigg(\frac{\partial}{\partial t}+\nu( k^2+k'^2)\bigg)\langle{\bf u}(k){\bf u}(k')\rangle=\langle{\bf u}{\bf u}{\bf u}\rangle.
\label{velocity second order correlation equation in Fourier space}
\end{eqnarray}
We can derive the third order correlation term, which needs the fourth order correlation:
\begin{eqnarray}
\bigg(\frac{\partial}{\partial t}+\nu( k^2+p^2+q^2)\bigg)\langle{\bf u}(k){\bf u}(p){\bf u}(q)\rangle=
\langle{\bf u}{\bf u}{\bf u}{\bf u}\rangle.
\label{velocity third order correlation equation in Fourier space}
\end{eqnarray}
It is known that the probability distribution of turbulent velocity is not far from the normal distribution. Then, the fourth order correlation term can be decomposed into the combination of the second order correlation terms(Quasi Normal approximation, \cite{1954RSPTA.247..163P}, \cite{1957RSPSA.239...16T}).
\begin{eqnarray}
\langle{\bf u}(k){\bf u}(p){\bf u}(q){\bf u}(r)\rangle\sim \sum\langle{\bf u}{\bf u}\rangle\langle {\bf u}{\bf u}\rangle.
\label{decomposition of fourth order temr}
\end{eqnarray}

\subsection{Eddy Damping coefficient}
However, \cite{1963JFM....16...33O} pointed out that Quasi Normal approximation could make the energy spectrum negative. Later \cite{1970JFM....41..363O} found that the decomposed value became too large when the fourth correlation was decomposed of the combination of second correlation terms. Orszag introduced eddy damping coefficient $\mu_{kpq}$($\sim 1/t$).
\begin{eqnarray}
\bigg(\frac{\partial}{\partial t}+\nu( k^2+p^2+q^2)+\mu_{kpq}\bigg)\langle{\bf u}(k){\bf u}(p){\bf u}(q)\rangle\nonumber\\
=\sum\langle{\bf u}{\bf u}\rangle\langle{\bf u}{\bf u}\rangle.
\label{EDQN Eddy eddy damping term 1}
\end{eqnarray}
Orszag suggested
\begin{eqnarray}
\mu_{kpq}=\mu_k+\mu_p+\mu_q,\quad\,\, \mu_k\sim[k^3E(k)]^{1/2}.
\label{EDQN Eddy eddy damping term form}
\end{eqnarray}
($\mu_{kpq}$ used in Eq.\ref{kinetic nu theta mu definition} is a little different from Orszag's one.) However, if energy drops faster than $k^{-3}$, eddy damping term($\sim t^{-1}$) decreases with `$k$'. This means the damping time of a smaller eddy can be larger than that of a larger eddy. To solve this problem, another modified representation was suggested by \cite{1978JMec...17..609L}:
\begin{eqnarray}
\mu_k\sim \bigg[\int^k_0E(p,t)\,dp\bigg]^{1/2}.
\label{EDQN Eddy eddy damping term form}
\end{eqnarray}
This is `$Eddy\,\,Damped\,\,Quasi\,\,Normal$' approximation.\\
Then we have,
\begin{eqnarray}
&&\bigg(\frac{\partial}{\partial t}+2\nu k^2\bigg)\langle u_iu_j \rangle_{k,\,t}=\nonumber\\
&&\int_0^td\tau \int_{k+p+q=0}e^{-[\mu_{kpq}+\nu(k^2+p^2+q^2)](t-\tau)}\sum \langle{\bf u}{\bf u}\rangle\langle{\bf u}{\bf u}\rangle dp\,dq.\nonumber\\
\label{EDQN Solution 1}
\end{eqnarray}
If time scale of $\sum \langle{\bf u}{\bf u}\rangle\langle{\bf u}{\bf u}\rangle$ is much larger than $[\mu_{kpq}+\nu(k^2+p^2+q^2]^{-1}$, markovianization makes the equation much simpler.
\begin{eqnarray}
\bigg(\frac{\partial}{\partial t}+2\nu k^2\bigg)\langle u_iu_j \rangle_{k,\,t}=\int_{k+p+q=0}\theta_{kpq}\sum \langle{\bf u}{\bf u}\rangle\langle{\bf u}{\bf u}\rangle dp\,dq.\\
\theta_{kpq}=\int_0^td\tau e^{-[\mu_{kpq}+\nu(k^2+p^2+q^2)](t-\tau)}\,d\tau\nonumber\\
\label{EDQNM Solution}
\end{eqnarray}
This is called `$Eddy\,\,Damped\,\,Quasi\,\,Normal\,\,Markovian$ approximation'(EDQNM, \cite{1976JFM....77..321P}, \cite{2001inma.book.....D}, \cite{2008tufl.book.....L}).

\bibliographystyle{mn2e}
\bibliography{bibdatabase}
\end{document}